\documentstyle[amsmath, amssymb, epsf, 11pt]{article}
\begin{document}

\date{\today }
\title{ {\bf Cosmology in three dimensions: steps towards the general
solution}}
\author{John D. Barrow${}^{1}$, Douglas J. Shaw${}^{1}$ and Christos G.
Tsagas${}^{2}$ \\
%EndAName
{\small ${}^1$DAMTP, Centre for Mathematical Sciences, University of
Cambridge}\\
{\small Wilberforce Road, Cambridge CB3 0WA, UK}\\
${}^2${\small Section of Astrophysics, Astronomy and Mechanics, Department
of Physics}\\
{\small Aristotle University of Thessaloniki, Thessaloniki 54124, Greece}}
\maketitle

\begin{abstract}
We use covariant and first-order formalism techniques to study the
properties of general relativistic cosmology in three dimensions.
The covariant approach provides an irreducible decomposition of the
relativistic equations, which allows for a mathematically compact
and physically transparent description of the 3-dimensional
spacetimes. Using this information we review the features of
homogeneous and isotropic 3-d cosmologies, provide a number of new
solutions and study gauge invariant perturbations around them. The
first-order formalism is then used to provide a detailed study of
the most general 3-d spacetimes containing perfect-fluid matter.
Assuming the material content to be dust with comoving spatial
2-velocities, we find the general solution of the Einstein equations
with non-zero (and zero) cosmological constant and generalise known
solutions of Kriele and the 3-d counterparts of the Szekeres
solutions. In the case of a non-comoving dust fluid we find the
general solution in the case of one non-zero fluid velocity
component. We consider the asymptotic behaviour of the families of
3-d cosmologies with rotation and shear and analyse their singular
structure. We also provide the general solution for cosmologies with
one spacelike Killing vector, find solutions for cosmologies
containing scalar fields and identify all the PP-wave 2+1
spacetimes.
\end{abstract}

\date{ }

\section{Introduction}

%%%%%%%%%%%%%%%%%%%%%%
General relativity in three spacetime dimensions is known to possess a
number of special simplifying features: there are no gravitational waves, no
black holes in the absence of a negative cosmological constant, the Weyl
curvature is identically zero, and the weak-field limit of the theory does
not correspond to Newtonian gravity in two space dimensions~\cite{GAK}-\cite%
{teit}. The theory is therefore considerably 'smaller' than general
relativity spacetimes with four (or more) dimensions, and the strong-energy
condition that creates geodesic focussing does not depend on the density of
the material sources. These simplifying features mean that considerable
progress can be made in the search for the general cosmological solution of
the three-dimensional Einstein equations. In an $(N+1)$-dimensional
spacetime the number of independently arbitrary $N$-dimensional functions of
the space coordinates that are needed to specify the Cauchy data for the
general cosmological problem on a spacelike hypersurface in vacuum is $%
(N+1)(N-2)$; in the presence of a general (non-comoving) perfect fluid it is
$N^{2}-1;$ and for a comoving perfect fluid it is $N^{2}-N-1$~\cite{BBL}.
Thus, in the $N=2$ case, we see that the number reduces to zero for the
vacuum solution (reflecting the absence of free gravitational fields in
vacuum), reduces to one arbitrary spatial function in the comoving
perfect-fluid case ,and to three arbitrary spatial functions for a perfect
fluid.

In this paper we will set up the general cosmological problem in
three-dimensional spacetimes and find the general solution of the field
equations in the case of comoving pressure-free matter, with and without a
cosmological constant, $\Lambda $. We go on to find solutions for the case
of non-comoving dust and classify the singularities and asymptotic
behaviours that arise in both cases with and without a cosmological
constant. The relative tractability of the general cosmological problem in
2+1 dimensions allows us to go some way towards finding a general solution
of the Einstein equations and we are able to isolate those features which
prevent a full solution being found. In particular, we are able to find and
classify the solutions for dust containing one of the (two possible)
non-zero spatial 2-velocity components.

There have been several past investigations of the structure of 2+1
dimensional general relativity and studies of the properties of particular
solutions with high symmetry (see~\cite{GAK}-\cite{D}, and~\cite{Clema}-\cite%
{Ida}). Important motivations for these studies were provided by the
astrophysical interest in the possible observational signatures of cosmic
strings and domain walls in the universe \cite{vil}-\cite{his}. Higher-order
curvature contributions were discussed in~\cite{BBL}, together with the
special features of the Newtonian-relativistic correspondence in general
relativity and related theories, while the study of quantum gravity is
reviewed in~\cite{carl}. Cosmological solutions and singularities were
discussed in~\cite{BBL} and~\cite{collas,gar1}, while gravitational collapse
of spherically symmetric dust clouds have been considered in~\cite{Kriele}-%
\cite{gutti}.

The outline of this paper is as follows. In section 2 we define the 3-d
Einstein equations and our notations. Section 3 introduces the 2+1 covariant
formalism and the general kinematics of 3-d spacetimes, identifying the
special features that arise from the lower dimensions and from the vanishing
of the Weyl curvature. These include the key role of the isotropic pressure
as the sole contributor to the gravitational mass of the system and the fact
that vorticity never increases with time. In section 4 we give a number of
new cosmological solutions, review the characteristics of the homogeneous
and isotropic models, including those that are singularity-free, and provide
the generalisation of the G\"{o}del universe to 3 dimensions. We also
consider linear perturbations around the 3-d analogues of the
`dust'-dominated FRW models and find them to be (neutrally) stable. In
section 5 we employ Witten's first-order formalism~\cite{witten} to
formulate the equations governing the most general 3-d cosmological
spacetime metric containing perfect-fluid matter. Then, in section 6 we
specialise the matter source to pressure-free dust with non-zero $\Lambda $
and comoving 2-velocities and find the general solution of the field
equations. These fall into three classes, one of which generalises the
solution of Kriele~\cite{Kriele} to $\Lambda \neq 0$, while another is the
generalisation of the Szekeres metric with nonzero $\Lambda $~to 2+1
dimensions \cite{szek, BSS}. Section 7 considers the most general dust
cosmologies with non-comoving velocities and finds various new classes of
solutions. We study the asymptotic behaviour of these solutions and analyse
in detail the structure of their spacelike and timelike singularities. Also,
by means of a number of examples, we illustrate the wide range of possible
behaviours in the presence of vorticity and shear. The same section also
introduces a transformation that generates exact solutions with nonzero
cosmological constant from those with vanishing $\Lambda $. Finally, in
section 8 we look at the case of a pure scalar field, provide the general
solution of Einstein's equations with one spacelike Killing vector, and
identify all the 2+1 PP-wave spacetimes. Our results are summarised and
discussed in section 9.

\section{Einstein's equations}

%%%%%%%%%%%%%%%%%%%%%%%%%%%%%%
It has long been known that general relativity, as a theory of a Riemannian
spacetime, can be based on a small number of generally accepted postulates
which are independent of the spacetime dimensions~\cite{V}-\cite{W}.
Assuming non-zero cosmological constant, these postulates demand that the
field equations take the form
\begin{equation}
R_{ab}-{\textstyle{\frac{1}{2}}}Rg_{ab}+\Lambda g_{ab}=\kappa T_{ab}\,,
\label{EFE1}
\end{equation}%
where $R_{ab}$ is the Ricci tensor, with $R=R_{a}{}^{a}$, $T_{ab}$ is the
energy-momentum tensor of the matter generating the metric field $g_{ab}$, $%
\Lambda $ is the cosmological constant and $\kappa $ is a dimensional
coupling constant.\footnote{%
Throughout this paper Latin indices run between 0 and 2 and Greek take the
values 1 and 2. Also, the three-dimensional coupling constant $\kappa $ is
measured in units of mass${}^{-1}$ and therefore defines a natural mass unit~%
\cite{GAK}.} When dealing with 3-dimensional spacetimes $g_{ab}g^{ab}=\delta
_{a}{}^{a}=3$. In this geometrical environment $R=-2\kappa T+6\Lambda $,
with $T=T_{a}{}^{a}$, and Einstein's equations become (e.g.~see~\cite%
{GAK,BBL})
\begin{equation}
R_{ab}=\kappa (T_{ab}-Tg_{ab})+2\Lambda g_{ab}\,.  \label{EFE2}
\end{equation}

The spacetime geometry is determined by the Riemann curvature tensor $%
R_{abcd}$. In three dimensions the latter has six independent components,
exactly as many as the associated Ricci tensor. This means that the
spacetime geometry can be expressed solely in terms of the Ricci curvature,
namely that~\cite{E,S}
\begin{equation}
R_{abcd}=g_{ac}R_{bd}+g_{bd}R_{ac}-g_{bc}R_{ad}-g_{ad}R_{bc}-{\textstyle{%
\frac{1}{2}}}R(g_{ac}g_{bd}-g_{ad}g_{bc})\,.  \label{Riemann}
\end{equation}%
As a result, the 3-dimensional Weyl tensor vanishes identically and the
gravitational field has no dynamical degrees of freedom. The spacetime
curvature is completely determined by the local matter distribution and the
theory is Machian.

\section{Covariant decomposition}

%%%%%%%%%%%%%%%%%%%%%%%%%%%

\subsection{Observers}

%%%%%%%%%%%%%%%%%%%%%%
In analogy to the standard 3+1 covariant approach to general relativity
introduced by Ehlers \cite{ehlers} and elaborated by Ellis (e.g.~see~\cite%
{EvE} for a recent review), we introduce a family of timelike (fundamental)
observers with worldlines tangent to the 3-velocity field $u_{a}$. The
latter determines the time direction and is normalised so that $u_{a}u^{a}=-1
$. The 2-dimensional space is defined by projecting orthogonal to $u_{a}$ by
means of the projection tensor
\begin{equation}
h_{ab}=g_{ab}+u_{a}u_{b}\,,  \label{hab}
\end{equation}%
where $h_{ab}=h_{(ab)}$, $h_{ab}u^{b}=0$, $h_{ab}h^{b}{}_{c}=h_{ac}$ and $%
h_{a}{}^{a}=2$. Using $h_{ab}$ one also defines the covariant derivative
operator of the 2-d space as ${\rm D}_{a}=h_{a}{}^{b}\nabla _{b}$.

The irreducible kinematical variables, which describe the motion of the
above defined observers in a invariant way, are obtained by decomposing the
covariant derivative of the 3-velocity. This splitting gives
\begin{equation}
\nabla _{b}u_{a}=\sigma _{ab}+\omega _{ab}+{\textstyle{\frac{1}{2}}}\Theta
h_{ab}-\dot{u}_{a}u_{b}\,,  \label{nablau}
\end{equation}%
where $\sigma _{ab}={\rm D}_{\langle b}u_{a\rangle }$ is the shear, $\omega
_{ab}={\rm D}_{[b}u_{a]}$ is the vorticity, $\Theta ={\rm D}^{a}u_{a}=\nabla
^{a}u_{a}$ is the area expansion (or contraction) scalar and $\dot{u}%
_{a}=u^{b}\nabla _{b}u_{a}$ the 3-acceleration. Therefore, $\sigma
_{ab}u^{b}=0=\omega _{ab}u^{b}=\dot{u}_{a}u^{a}$ by construction.

The tensor ${\rm D}_{b}u_{a}\equiv h_{b}{}^{d}h_{a}{}^{c}\nabla
_{d}u_{c}=\sigma _{ab}+\omega _{ab}+(\Theta /2)h_{ab}$ describes changes in
the relative position of the worldlines of two neighbouring observers. When
the latter follow the motion of a fluid, the effect of $\Theta $ is to
change the area of a given fluid element, without causing rotation or shape
distortion. This scalar also defines the average scale factor, $a,$ by
\begin{equation}
\frac{\dot{a}}{a}={\textstyle{\frac{1}{2}}}\Theta \,.  \label{a}
\end{equation}%
The shear monitors distortions in the element's shape that leave the area
unaffected, while $\omega _{ab}$ describes changes in its orientation under
constant area and shape. The symmetric and trace-free nature of $\sigma
_{ab} $ ensures that it has only two independent components, while the
antisymmetry of $\omega _{ab}$ guarantees that the vorticity tensor is
determined by a single component. In other words, the shear and the
vorticity correspond to a vector and a scalar respectively. The latter
reflects the fact that the rotational axis has been reduced to a point.
Defining $\epsilon _{ab}=\epsilon _{\lbrack ab]}$ as the 2-dimensional
permutation tensor, with $\epsilon _{ab}u^{b}=0$, the vorticity scalar is
\begin{equation}
\omega ={\textstyle{\frac{1}{2}}}\epsilon _{ab}\omega ^{ab}\,,  \label{omega}
\end{equation}%
with $\omega _{ab}=\omega \epsilon _{ab}$. Note that $\epsilon _{ab}=\eta
_{abc}u^{c}$ by definition, where $\eta _{abc}$ is the 3-d totally
antisymmetric alternating tensor. The latter satisfies the condition $\eta
_{abc}\eta ^{dqs}=3!\delta _{\lbrack a}{}^{d}\delta _{b}{}^{q}\delta
_{c]}{}^{s}$, which ensures that $\epsilon _{ab}\epsilon
^{cd}=2h_{[a}{}^{c}h_{b]}{}^{d}$.

\subsection{Matter fields}

%%%%%%%%%%%%%%%%%%%%%%%
Suppose that the matter that sources the 3-dimensional metric field is a
perfect fluid. Then, relative to an observer moving with 3-velocity $u_{a}$,
the energy-momentum tensor of the material component takes the form
\begin{equation}
T_{ab}=\rho u_{a}u_{b}+ph_{ab}\,,  \label{Tab}
\end{equation}%
where $\rho $ is the energy density, $p$ is pressure and its trace is $%
T=2p-\rho $. Substituting the above into the Einstein field equations (\ref%
{EFE2}) the latter read
\begin{equation}
R_{ab}=2(\kappa p-\Lambda )u_{a}u_{b}+[\kappa (\mu -p)+2\Lambda ]h_{ab}\,,
\label{pfRab}
\end{equation}%
with trace $R=2[\kappa (\mu -2p)+3\Lambda ]$. The above also provides the
following auxiliary relations
\begin{equation}
R_{ab}u^{a}u^{b}=2(\kappa p-\Lambda )\,,\hspace{10mm}%
h_{a}{}^{c}h_{b}{}^{d}R_{cd}=[\kappa (\rho -p)+2\Lambda ]h_{ab}\hspace{5mm}%
{\rm and}\hspace{5mm}h_{a}{}^{b}R_{bc}u^{c}=0\,  \label{EFE3}
\end{equation}%
which will prove useful later.

The twice-contracted Bianchi identities imply that $\nabla ^{b}G_{ab}=0$ and
consequently the condition $\nabla ^{b}T_{ab}=0$. The timelike and spacelike
parts of the latter lead to the 3-d fluid conservation laws. These are
\begin{equation}
\dot{\rho}=-\Theta (\rho +p)\,,  \label{edc}
\end{equation}%
for the energy density of the fluid, and
\begin{equation}
(\rho +p)\dot{u}_{a}=-{\rm D}_{a}p  \label{mdc}
\end{equation}%
for its momentum density. The above ensure that the conservation
laws of a perfect fluid have the same functional form as their
4-dimensional counterparts (compare to Eqs.~(37), (38)
of~\cite{EvE}).

The nature of the medium is determined by its equation of state.
Here we will only consider barotropic fluids with $p=w\rho$, where
$w$ represents the barotropic index. When $w=0$ we are dealing with
pressure-free dust, while isotropic radiation has $p=\rho/2$ and
corresponds to $w=1/2$~\cite{GAK}.

\subsection{Spatial curvature}

%%%%%%%%%%%%%%%%%%%%%%%%%%%
The intrinsic curvature of the 2-dimensional space orthogonal to $u_{a}$ is
determined by the associated Riemann tensor. In analogy with its standard
3-d counterpart (see Eq.~(77) in~\cite{El}), the latter is defined by
\begin{equation}
{\cal R}%
_{abcd}=h_{a}{}^{q}h_{b}{}^{s}h_{c}{}^{f}h_{d}{}^{p}R_{qsfp}-v_{ac}v_{bd}+v_{ad}v_{bc}\,,
\label{2Riemann1}
\end{equation}%
where
\begin{equation}
v_{ab}={\rm D}_{b}u_{a}=\sigma _{ab}+\omega _{ab}+{\textstyle{\frac{1}{2}}}%
\Theta h_{ab}\,,  \label{vab}
\end{equation}%
is the relative position vector. Note that $v_{ab}$ characterises the
extrinsic curvature (i.e.~the second fundamental form) of the space.

Starting from Eq.~(\ref{2Riemann1}), assuming perfect-fluid matter and using
expressions (\ref{Riemann}), (\ref{pfRab}), (\ref{EFE3}) and (\ref{vab}),
the Riemann tensor of the 2-d (spatial) sections reads
\begin{eqnarray}
{\cal R}_{abcd} &=&\left( \kappa \rho -{\textstyle{\frac{1}{4}}}\Theta
^{2}+\Lambda \right) (h_{ac}h_{bd}-h_{ad}h_{bc})-(\sigma _{ac}+\omega
_{ac})(\sigma _{bd}+\omega _{bd})+(\sigma _{ad}+\omega _{ad})(\sigma
_{bc}+\omega _{bc})  \nonumber \\
&{}&-{\textstyle{\frac{1}{2}}}\Theta \left[ (\sigma _{ac}+\omega
_{ac})h_{bd}+h_{ac}(\sigma _{bd}+\omega _{bd})-(\sigma _{ad}+\omega
_{ad})h_{bc}-h_{ad}(\sigma _{bc}+\omega _{bc})\right] \,,  \label{2Riemann2}
\end{eqnarray}%
with ${\cal R}_{abcd}={\cal R}_{[ab][cd]}$. In agreement with standard 3+1
gravity, the isotropic pressure does not contribute to the curvature of the
space orthogonal to $u_{a}$. Also, in the absence of anisotropy (i.e.~when $%
\sigma _{ab}$ and $\omega _{ab}$ vanish) the above reduces to
\begin{equation}
{\cal R}_{abcd}=\left( \kappa \rho -{\textstyle{\frac{1}{4}}}\Theta
^{2}+\Lambda \right) (h_{ac}h_{bd}-h_{ad}h_{bc})\,.  \label{iso2Riemann}
\end{equation}

Defining ${\cal R}_{ab}={\cal R}^{c}{}_{acb}$ as our local 2-D Ricci tensor,
we may contract expression (\ref{2Riemann2}) to obtain the following
3-dimensional analogue of the Gauss-Codacci formula (see Eq.~(54) in~\cite%
{EvE})
\begin{equation}
{\cal R}_{ab}=\left( \kappa \rho -{\textstyle{\frac{1}{4}}}\Theta
^{2}+\sigma ^{2}-\omega ^{2}+\Lambda \right) h_{ab}\,,  \label{cRab}
\end{equation}%
which here holds for perfect-fluid matter. In deriving the above we used the
results $\omega _{c[a}\sigma ^{c}{}_{b]}=0$ and $\sigma _{c\langle a}\sigma
^{c}{}_{b\rangle }=0=\omega _{c\langle a}\omega ^{c}{}_{b\rangle }$. The
former holds because $\omega _{12}(\sigma ^{1}{}_{1}+\sigma ^{2}{}_{2})=0$
(i.e.~the single independent component vanishes due to the trace-free nature
of the shear). Similarly, the two independent components of $\sigma
_{c\langle a}\sigma ^{c}{}_{b\rangle }$ are also identically zero. Last, the
result $\omega _{c\langle a}\omega ^{c}{}_{b\rangle }=0$ is guaranteed by
the relation $\omega _{ab}=\omega \epsilon _{ab}$ and the properties of $%
\epsilon _{ab}$ (see section 3.1). The absence of a skew and also of a
symmetric and trace-free part from Eq.~(\ref{cRab}) agrees with symmetries
of the Riemann tensor in 2-d spaces (e.g.~see~\cite{GAK}).

Finally, the trace of (\ref{cRab}) leads to the curvature scalar of the
spatial sections, which may also be seen as the generalised Friedmann
equation for three-dimensional spacetimes (compare to Eq.~(55) of~\cite{EvE}%
)
\begin{equation}
{\cal R}\equiv {\cal R}^{a}{}_{a}=2\left( \kappa \rho -{\textstyle{\frac{1}{4%
}}}\Theta ^{2}+\sigma ^{2}-\omega ^{2}+\Lambda \right) \,.  \label{cR}
\end{equation}

\subsection{Kinematics}

%%%%%%%%%%%%%%%%%%%%
The functional form of the Ricci identity is independent of dimension. Thus,
when applied to the 3-velocity vector $u_{a},$ the Ricci identity reads
\begin{equation}
\nabla _{a}\nabla _{b}u_{c}-\nabla _{b}\nabla _{a}u_{c}=R_{abcd}u^{d}\,,
\label{Ricci}
\end{equation}%
where $R_{abcd}$ is the Riemann tensor of the 3-d spacetime (see expression (%
\ref{Riemann})). Contracting the above along $u_{b}$, employing Eqs.~(\ref%
{Riemann}), (\ref{EFE3}), and then taking the trace of the resulting
expression we obtain the 3-d analogue of Raychaudhuri's equation
\begin{equation}
\dot{\Theta}=-{\textstyle{\frac{1}{2}}}\Theta ^{2}-2\kappa p-2(\sigma
^{2}-\omega ^{2})+{\rm D}^{a}\dot{u}_{a}+\dot{u}^{a}\dot{u}_{a}+2\Lambda \,,
\label{Ray}
\end{equation}%
with $2\sigma ^{2}\equiv \sigma _{ab}\sigma ^{ab}$ and $2\omega ^{2}\equiv
\omega _{ab}\omega ^{ab}$. The above is the key equation of gravitational
attraction, as it monitors the average separation between neighbouring
particle worldlines. The most important difference between (\ref{Ray}) and
its 4-d counterpart (see Eq.~(29) in~\cite{EvE}) is that here only the fluid
pressure contributes to the gravitational mass of the medium: the density $%
\rho $ does not contribute. This unusual feature consists a major departure
from standard gravity. One consequence is the existence of homogeneous and
isotropic static 3-dimensional models with dust and zero cosmological
constant (see solution (\ref{adust}) below).

The symmetric and trace-free part of the contracted Ricci identity, together
with the auxiliary relations (\ref{EFE3}) leads to the propagation formula
for the shear in three dimensions:
\begin{equation}
h_{a}{}^{c}h_{b}{}^{d}\dot{\sigma}_{cd}=-\Theta \sigma _{ab}+{\rm D}%
_{\langle b}\dot{u}_{a\rangle }+\dot{u}_{\langle a}\dot{u}_{b\rangle }\,.
\label{sigmadot1}
\end{equation}%
Relative to the 4-dimensional case (see Eq.~(30) in~\cite{EvE}), we
notice the absence of the electric Weyl component from the
right-hand side of this formula. This reflects the vanishing of the
free gravitational field in 3-d gravity. Contracting this expression
with $\sigma _{ab}$ we obtain the propagation formula for the shear
magnitude
\begin{equation}
\left( \sigma ^{2}\right) ^{\cdot }=-2\Theta \sigma ^{2}+\sigma ^{ab}{\rm D}%
_{\langle b}\dot{u}_{a\rangle }+\dot{u}_{\langle a}\dot{u}_{b\rangle }\sigma
^{ab}\,,  \label{sigma2dot}
\end{equation}%
where $2\sigma ^{2}=\sigma _{ab}\sigma ^{ab}$ by definition. In the absence
of fluid accelerations the 2nd and 3rd terms on the right-hand side vanish
and the equation integrates to give $\sigma ^{2}\propto a^{-4}$.

Similarly, the contracted antisymmetric component of (\ref{Ricci}) gives the
3-d counterpart of the vorticity propagation equation. The latter reads
\begin{equation}
h_{a}{}^{c}h_{b}{}^{d}\dot{\omega}_{cd}=-\Theta \omega _{ab}+{\rm D}_{[b}%
\dot{u}_{a]}\,.  \label{omegadot}
\end{equation}%
Comparing to Eq.~(31) of~\cite{EvE} we note the absence of the $\omega
_{c[a}\sigma ^{c}{}_{b]}$ term. This is so because $\omega _{c[a}\sigma
^{c}{}_{b]}=0$ in 3-d (see also above). Also, contracted with $\omega _{ab}$%
, and recalling that $2\omega ^{2}\equiv \omega _{ab}\omega ^{ab}$ Eq.~(\ref%
{omegadot}) gives the scalar vorticity propagation formula:
\begin{equation}
\left( \omega ^{2}\right) ^{\cdot }=-2\Theta \omega ^{2}+\omega ^{ab}{\rm D}%
_{[b}\dot{u}_{a]}\,.  \label{omega2dot}
\end{equation}%
Again, in the absence of accelerations when the pressure vanishes this
equation integrates to give $\omega ^{2}\propto a^{-4}$ as expected by the
conservation of angular momentum. For a barotropic perfect fluid with $%
p=p(\rho )$ and a constant barotropic index $w=p/\rho $, Eqs.~(\ref{edc}), (%
\ref{mdc}) combine with the commutation law ${\rm D}_{[a}{\rm D}_{b]}f=-\dot{%
f}\omega _{ab}$ to recast the vorticity propagation formula as
\begin{equation}
h_{a}{}^{c}h_{b}{}^{d}\dot{\omega}_{cd}=-(1-c_{{\rm s}}^{2})\Theta \omega
_{ab}\,,  \label{omegadot1}
\end{equation}%
where $c_{{\rm s}}^{2}={\rm d}p/{\rm d}\rho =w$ is the square of the
adiabatic sound speed. Therefore, the expansion decreases the
vorticity of a 2-dimensional space unless the barotropic fluid has a
stiff equation of state (i.e.~when $c_{{\rm s}}^{2}=1=w$) and for
general perfect fluids we have $\omega ^{2}\propto a^{-2(1-w)}$.
Recall that, in standard general relativity, vorticity increases
when $w>2/3$~\cite{jdbv}.

The simultaneous effects of shear and vorticity on a cosmology with
negligible accelerations can be evaluated from these simple relations. For
all equations of state we have $\sigma ^{2}\propto a^{-4}$ but the
centrifugal energy depends on the equation of state since $\omega
^{2}\propto a^{-4(1-w)}$. Hence, the ratio of the distortion energy density
to the centrifugal energy density is

\[
\frac{\sigma ^{2}}{\omega ^{2}}\propto a^{-2w}
\]%
and the shear always dominates as $a\rightarrow 0$ when $p>0$ but the
vorticity always dominates as $a\rightarrow \infty $. The presence of fluid
acceleration can modify this behaviour; an example with $p=0$ will be given
below.

\section{Cosmology in 2+1 dimensional spacetimes}

%%%%%%%%%%%%%%%%%%%%%%%%%%%%%%%%%%%%%%%

\subsection{Homogeneous and isotropic spacetimes}

%%%%%%%%%%%%%%%%%%%%%%%%%%%%%%%%%%%%%%%%%%%%%%%%%
Spatial homogeneity and isotropy means that we can always choose a
coordinate system such as the line element of the 3-dimensional spacetime
takes the Friedmann-Robertson-Walker (FRW) form
\begin{equation}
{\rm d}s^{2}=-{\rm d}t^{2}+a^{2}h_{\alpha \beta }{\rm d}x^{\alpha
}{\rm d} x^{\beta }\,,  \label{2RW}
\end{equation}%
where the scale factor $a=a(t)$ completely determines the time
evolution of the model. The form also represents the metric of the
3-dimensional analogue of the familiar FRW cosmologies. The
kinematics of (\ref{2RW}) is monitored via one propagation and one
constraint equation (see expressions (\ref{Ray}) and (\ref{cR})).
Assuming a barotropic fluid with with constant barotropic index,
setting $\Lambda =0$ and recalling that $\Theta =2\dot{a}/a$, these
formulae recast into
\begin{equation}
\frac{\ddot{a}}{a}=-2\kappa w\rho \hspace{15mm}{\rm and}\hspace{15mm}\left(
\frac{\dot{a}}{a}\right) ^{2}=\kappa \rho -\frac{k}{a^{2}}\,,  \label{kin}
\end{equation}%
respectively ($k=0,\,\pm 1$ is the curvature index of the spatial
sections). The matter component obeys the conservation law
(\ref{edc}), which accepts the solution
\begin{equation}
\rho =\rho _{0}\left( \frac{a_{0}}{a}\right) ^{2(1+w)}\,,  \label{matter}
\end{equation}%
with $a_{0}$ constant.

For dust, $w=0,$ and one immediately finds (see Eq.~(\ref{kin}a)) that $%
a\propto t$ for all signs of $k$. More specifically, expression (\ref{matter}%
) gives $\rho \propto a^{-2}$, which substituted into Eq.~(\ref{kin}b) leads
to
\begin{equation}
a=\left( \kappa \rho _{0}a_{0}^{2}-k\right) ^{1/2}t+a_{0}\,,  \label{adust}
\end{equation}%
where the product $\rho _{0}a_{0}^{2}$ is proportional to the total mass of
the model. Therefore, the scale factor of a dust-dominated, 3-dimensional,
FRW universe evolves linearly with time, irrespective of its spatial
curvature and no collapse to a final singularity occurs. Note that when $%
\kappa \rho _{0}a_{0}^{2}=k$ we obtain a static solution with $a=a_{0}$~\cite%
{GAK}. Unlike its 4-d counterparts, this static universe has zero
cosmological constant (see also section 5 above).

When the matter content is in the form of black-body radiation, the
barotropic index is $w=1/2$. Then, expression (\ref{matter}) gives $\rho
\propto a^{-3}$ and the system (\ref{kin}) has the solution
\begin{equation}
a=\int \sqrt{\frac{\kappa \rho _{0}a_{0}^{3}-ka}{a}}\,{\rm d}t+a_{0}\,.
\label{arad}
\end{equation}%
For $k=0$ the above reduces to $a\propto t^{2/3}$, which coincides with the
scale-factor evolution in standard 3+1 dust-dominated FRW universes.

One can study perturbations around the above given homogeneous and isotropic
solutions by introducing the covariantly defined variables ${\cal D}%
_{a}=(a/\rho ){\rm D}_{a}\rho $ and ${\cal Z}_{a}=a{\rm D}_{a}\Theta
$. The former describes variations in the matter density and the
latter in the area expansion as measured by a pair of neighbouring
observers~\cite{EB}. Also, both variables vanish in the spatially
homogeneous background and therefore satisfy the gauge-invariance
criterion. In the case of a pressureless fluid (i.e.~for
$w=0=c_{{\rm s}}^{2}$) the linear evolution of inhomogeneities is
monitored by
\begin{equation}
\dot{{\cal D}}=-{\cal Z}\hspace{10mm}{\rm and}\hspace{10mm}\dot{{\cal Z}}%
=-\Theta {\cal Z}\,,  \label{cDcZdot}
\end{equation}%
on all scales. Then, using solution (\ref{adust}) and setting ${\cal D}%
_{0}=(\alpha t_{0}+a_{0}){\cal Z}_{0}/\alpha $ initially, we find that the
density gradient decays as
\begin{equation}
{\cal D}={\cal D}_{0}\left( \frac{\alpha t_{0}+a_{0}}{\alpha t+a_{0}}\right)
\,,  \label{dcD}
\end{equation}%
with $\alpha =\sqrt{\kappa \rho _{0}a_{0}^{2}-k}$ and $k=0,\pm 1$ (see (\ref%
{adust})). For general initial conditions, on the other hand, it is
straightforward to show that ${\cal D}\rightarrow {\cal D}_{0}-(\alpha
t_{0}+a_{0}){\cal Z}_{0}/\alpha $ at late times. Recalling that for zero
pressure shear and vorticity perturbations decay in time (see Eqs.~(\ref%
{sigmadot1}) and (\ref{omegadot})), we conclude that in the absence of
pressure the 3-d analogues of the FRW cosmologies are either stable or
neutrally stable. This behaviour is very different from that of conventional
FRW models, all of which are unstable to density perturbations, and reflects
the fact that in three dimensions the gravitational mass of a pressure-free
medium vanishes (see Eq.~(\ref{Ray})) and spherical regions of all
curvatures asymptote to $a\rightarrow t$ as $t\rightarrow \infty $. The
immediate consequence is the absence of linear Jeans-type instabilities in
these models.

\subsection{Homogeneous and anisotropic spacetimes}

%%%%%%%%%%%%%%%%%%%%%%%%%%%%%%%%%%%%%%%%%%%%%%%%%%%
The simplest 3-d line element describing a homogeneous and anisotropic
spacetime has the following Bianchi~I-type form
\begin{equation}
{\rm d}s^{2}=-{\rm d}t^{2}+A^{2}{\rm d}x^{2}+B^{2}{\rm d}y^{2}\,,
\label{2BianchiI}
\end{equation}%
where $A=A(t)$ and $B=B(t)$ are the two individual scale factors. When $%
\Lambda =0$, the spatial flatness of the above metric means that the
3-d analogue of the Bianchi~I universe is covariantly described by
the following set of propagation equations
\begin{equation}
\dot{\rho}=-(1+w)\Theta \rho \,,\hspace{15mm}\dot{\Theta}=-{\textstyle{\frac{%
1}{2}}}\Theta ^{2}-2\kappa w\rho -2\sigma ^{2}\,,\hspace{15mm}\dot{\sigma}%
=-\Theta \sigma \,,\text{ \ \ \ \ \ \ \ \ }\omega \equiv 0  \label{2BIprop}
\end{equation}%
supplemented by the constraint
\begin{equation}
{\textstyle{\frac{1}{4}}}\Theta ^{2}=\kappa \rho +\sigma ^{2}\,.
\label{2Bcon}
\end{equation}%
The above are obtained from Eqs.~(\ref{edc}), (\ref{Ray}),
(\ref{sigma2dot}) and (\ref{cR}), respectively, after dropping their
inhomogeneous terms and assuming spatial flatness together with zero
vorticity. Setting $\Theta =2\dot{a}/a$,
where $a$ represents the geometric-mean scale factor, expressions (\ref%
{2BIprop}a) and (\ref{2BIprop}c) lead to the evolution laws
\begin{equation}
\rho =\rho _{0}\left( \frac{a_{0}}{a}\right) ^{2(1+w)} \hspace{10mm}
{\rm and} \hspace{10mm} \sigma =\sigma _{0}\left(
\frac{a_{0}}{a}\right) ^{2}\,, \label{2BIev}
\end{equation}%
for the matter energy density and the shear respectively.
Substituting these results to Eq.~(\ref{2Bcon}) and assuming a
spacetime filled with pressure-free dust we arrive at
\begin{equation}
a=a_{0}\sqrt{\kappa \rho _{0}t^{2}+2\sigma _{0}t}\,.
\label{2BIadust}
\end{equation}%
As expected, for $\sigma _{0}=0$ the above result reduces to its
isotropic counterpart (see Eq.~(\ref{adust})). The same also happens
at late times, when the shear contribution to solution
(\ref{2BIadust}) becomes negligible.

Note that one can use the evolution law of the average scale factor
to obtain those of the individual ones. Assuming dust, recalling
that
\begin{equation}
\frac{\dot{A}}{A}+\frac{\dot{B}}{B}=\Theta \hspace{15mm}{\rm
and}\hspace{15mm} \frac{\dot{A}}{A}-\frac{\dot{B}}{B}=2\sigma \,,
\label{ABprop}
\end{equation}%
and using results (\ref{2BIev}c) and (\ref{2BIadust}) we arrive at
\begin{equation}
A=A_{0}\left( \frac{t}{t_{0}}\right) \hspace{15mm}{\rm and}\hspace{15mm}%
B=B_{0}\left( \frac{\kappa \rho _{0}t+2\sigma _{0}}{\kappa \rho
_{0}t_{0}+2\sigma _{0}}\right) \,.  \label{ABdust}
\end{equation}%
In the vacuum case ($\rho _{0}\equiv 0$) the metric is just a coordinate
transformation of flat spacetime rather than an anisotropic cosmology.

\subsection{Rotating spacetimes}

%%%%%%%%%%%%%%%%%%%%%%%%%%%%%%%%
Consider a rotating three-dimensional spacetime with flat 2-d sections
filled with pressureless matter. Setting $\Lambda =0$ and assuming spatial
homogeneity, the time evolution of the model (at least locally) is monitored
by the following system of four propagation equations
\begin{equation}
\dot{\rho}=-\Theta \rho \,,\hspace{10mm}\dot{\Theta}=-{\textstyle{\frac{1}{2}%
}}\Theta ^{2}-2\sigma ^{2}+2\omega ^{2}\,,\hspace{10mm}\dot{\sigma}=-\Theta
\sigma \hspace{10mm}{\rm and}\hspace{10mm}\dot{\omega}=-\Theta \omega \,,
\label{rotprop}
\end{equation}%
constrained by
\begin{equation}
{\textstyle{\frac{1}{4}}}\Theta ^{2}=\kappa \rho +\sigma ^{2}-\omega ^{2}\,.
\label{rotcon}
\end{equation}%
Proceeding as before, we may use the area expansion scalar ($\Theta $) to
define an average scale factor ($a$) so that $\Theta =2\dot{a}/a$. Then,
expressions (\ref{rotprop}a), (\ref{rotprop}b) and (\ref{rotprop}c)
translate into
\begin{equation}
\rho =\rho _{0}\left( \frac{a_{0}}{a}\right) ^{2}\,,\hspace{10mm}\sigma
=\sigma _{0}\left( \frac{a_{0}}{a}\right) ^{2}\hspace{10mm}{\rm and}\hspace{%
10mm}\omega =\omega _{0}\left( \frac{a_{0}}{a}\right) ^{2}  \label{rotev}
\end{equation}%
respectively. Substituted into constraint (\ref{rotcon}) the above results
give
\begin{equation}
a=a_{0}\sqrt{\kappa \rho _{0}t^{2}+2t\sqrt{\sigma _{0}^{2}-\omega _{0}^{2}}}%
\,.  \label{rotadust}
\end{equation}%
Accordingly, despite the presence of nonzero shear and vorticity,
the average scale factor evolves as its homogeneous and isotropic
counterpart if $\sigma _{0}=\omega _{0}$. The situation in 3-d is
unusual in that the shear and vorticity scale as the same powers of
the scale factor and are both equally important at all times. In the
case of dust, the matter density also scales as $a^{-2}$ (see
(\ref{rotev})).

One can also obtain a 3-d G\"{o}del-type universe. The G\"{o}del spacetime
is a homogeneous spacetime and a rotating solution of Einstein's equations
which permits closed timelike curves~\cite{G, O}. Covariantly, G\"{o}del's
world is described by~\cite{HE, BT}
\begin{equation}
\Theta =0=\dot{u}_{a}=\sigma _{ab}\hspace{10mm}{\rm and}\hspace{10mm}\omega
_{ab}\neq 0\,.  \label{Godel}
\end{equation}%
Thus, with the exception of the vorticity, all the kinematical variables
vanish identically. Note that the overall homogeneity of the G\"{o}del
spacetime guarantees that the vorticity is a covariantly constant quantity
and that all the propagation equations reduce to constraints. Applying (\ref%
{Godel}) to three dimensions we arrive at the system
\begin{equation}
\dot{\rho}=0\,,\hspace{10mm}\dot{\omega}=0\,,\hspace{10mm}\kappa p-\omega
^{2}-\Lambda =0\hspace{10mm}{\rm and}\hspace{10mm}{\cal R}=2(\kappa \rho
-\omega ^{2}+\Lambda )\,.  \label{3DGodel}
\end{equation}%
which describes a 3-d G\"{o}del-type universe. Note that for
pressure-free matter the cosmological constant is necessarily
negative (i.e.~$\Lambda =-\omega ^{2}$ - see Eq.~(\ref{3DGodel}c)).
Also, unlike its standard 3+1 counterpart, the 3-dimensional
G\"{o}del-type spacetime can have non-vanishing spatial
Ricci curvature. This is guaranteed by (\ref{3DGodel}d). The empty G\"{o}%
del-type model with ${\cal R}=-4\omega ^{2}<0$ is equivalent to anti-de
Sitter (AdS) space. Rooman and Spindel, \cite{rooman}, considered a
one-parameter subset of these non-flat G\"{o}del-type solutions and showed
that they can be seen as arising from a directional squashing of the light
cones of AdS, which breaks the ${\frak so}(2,2)$ isometry of the AdS Killing
vectors into ${\frak so}(2,1)\times {\frak so}(2)$. It was shown in \cite%
{rooman} that all the non-empty G\"{o}del-type solutions considered
contained closed-timelike curves but they vanish in the AdS limit.

\subsection{Singularities}

%%%%%%%%%%%%%%%%%%%%%%%
The 3-d analogues of the standard singularity theorems are relatively
straightforward to deduce. The study of the Riemann and the Ricci curvature
shows that there are no Weyl curvature singularities and the analogue of the
strong-energy condition (i.e.~$R_{ab}u^{a}u^{b}\geq 0$) reduces to the
inequality $p\geq 0$ for a perfect fluid and to the positivity of the sum of
the principal pressures if they are anisotropic. Also, the form of the
Raychaudhuri equation (see (\ref{Ray})) guarantees that for geodesically
moving observers
\begin{equation}
\dot{\Theta}+{\textstyle{\frac{1}{2}}}\Theta ^{2}=-2\kappa p-2(\sigma
^{2}-\omega ^{2})+2\Lambda \,.  \label{geoRay1}
\end{equation}%
Therefore, for vanishing cosmological constant and in the absence of
rotation, an initially converging family of timelike worldlines will
focus (i.e.~$\Theta \rightarrow -\infty $). For non-zero vorticity,
however, this may not be necessarily the case. For example, applied
to a spatially homogeneous and rotating spacetime filled with an
isotropic radiation fluid, the above gives
\begin{equation}
\dot{\Theta}+{\textstyle{\frac{1}{2}}}\Theta ^{2}=-\kappa \rho _{0}\left(
\frac{a_{0}}{a}\right) ^{3}-2(\sigma _{0}^{2}-\omega _{0}^{2})\left( \frac{%
a_{0}}{a}\right) ^{4}\,.  \label{geoRay2}
\end{equation}%
Therefore, as $a\rightarrow 0$, the kinematical term dominates the
matter term in the right-hand side of the Raychaudhuri equation. In
this case, whether caustics will form or not depends entirely on the
balance between shear and rotation. When $\omega _{0}>\sigma _{0}$,
in particular, vorticity can stop an initially converging family of
worldlines from focusing.

There are also simple exact solutions that describe `bouncing' 3-d
cosmologies with $p<0$. Consider, for example, a perfect fluid with $p=-\rho
/2$. Then, assuming spatial homogeneity and isotropy, Eq.~(\ref{edc}) gives $%
\rho \propto a^{-1}$ and the Friedmann equation reduces to
\begin{equation}
\left( \frac{\dot{a}}{a}\right) ^{2}=\frac{M}{a}-\frac{k}{a^{2}}\,,
\end{equation}%
where $M>0$ is constant and $k$ is the spatial curvature index. The bouncing
solution is
\begin{equation}
a(t)=a_{{\rm min}}+{\textstyle{\frac{1}{4}}}M\left( t-t_{{\rm min}}\right)
^{2}\,,  \label{bounce1}
\end{equation}%
with $a_{{\rm min}}=k/M$. Accordingly, a non-singular minimum requires
positive curvature. Setting $k=+1$ and $t_{{\rm min}}=0$ the required
scale-factor evolution is represented by the parabola
\begin{equation}
a(t)=M^{-1}+{\textstyle{\frac{1}{4}}}Mt^{2}\,.  \label{bounce2}
\end{equation}

\section{First-Order Formalism}

In 2+1 dimensions it has been shown by Witten \cite{witten} that the
Palatini action for general relativity is equivalent to:

\begin{equation}
S_{g}=\int \tilde{\eta}^{abc}e_{a}^{I}\,{}^{3}F_{bcI}  \label{act}
\end{equation}

where $\tilde{\eta}^{abc}$ is the metric-independent alternating symbol in $%
2+1$ dimensions. Recall, our signature is $(-++)$ and we choose $\tilde{\eta}%
^{012}=-1\Rightarrow \tilde{\eta}_{012}=+1$; $e_{a}^{I}$ is a dreibein, and
in some sense should be viewed as the square root of the metric
\[
g_{ab}=e_{a}^{I}e_{b}^{J}\eta _{IJ},
\]%
where $\eta _{IJ}=diag(-1+1+1)$. We set

${}$%
\[
^{3}F_{abI}=2\partial _{\lbrack a}{}^{3}A_{b]}^{I}+\epsilon
_{IJK}{}^{3}A_{a}^{J}\,{}^{3}A_{b}^{K},
\]

where ${}^{3}A_{a}^{I}$ is an $SO(2,1)$ connection, and $\epsilon _{IJK}$ is
the standard alternating symbol. Here $a,b=0,1,2$ are spacetime indices,
whilst $I,J,K=\hat{0},\hat{1},\hat{2}$ are internal $SO(2,1)$ indices. We
raise and lower internal indices with $\eta _{IJ}$ and $\eta ^{IJ}$, so $%
\epsilon ^{IJK}=-\epsilon _{IJK}$, and we choose $\epsilon _{\hat{0}\hat{1}%
\hat{2}}=+1$. The action (\ref{act}) reduces to the standard second-order
gravitational action when ${}^{3}A_{a}^{I}$ is compatible with $e_{a}^{I}$
i.e.
\begin{equation}
{\cal D}_{[a}e_{b]I}:=\partial _{\lbrack a}e_{b]I}+\epsilon
_{IJK}\,{}^{3}A_{[a}^{J}e_{b]}^{K}=0.
\end{equation}%
We could also find this equation by varying the above action w.r.t. $%
{}^{3}A_{a}^{I}$. There is plenty of gauge freedom in this model, so we
firstly simplify matters by choosing to write the metric in synchronous
gauge as
\begin{equation}
{\rm d}s^{2}=g_{ab}{\rm d}x^{a}{\rm d}x^{b}=-{\rm d}t^{2}+h_{\alpha \beta }%
{\rm d}x^{\alpha }{\rm d}x^{\beta },
\end{equation}%
where $\alpha ,\beta =1,2$. In synchronous gauge: $e_{0}^{\hat{0}}=1$, $%
e_{0}^{i}=0$, $e_{\alpha }^{\hat{0}}=0$, where $i=\hat{1},\hat{2}$ are 2-d
internal indices. Throughout this section, we will use $a,b,c,d$ to
designate 3-d spacetime indices, and $I,J,K$ for 3-d internal indices. Lower
case Greek letters will be used for 2-d spacetime indices, and $i,j,k$ for
2-d internal ones. We define a projection operator onto the surfaces $t=const
$ by
\begin{equation}
h_{ab}=g_{ab}+n_{a}n_{b}
\end{equation}%
where $n_{a}=(1,0,0)$ and $h_{ab}=h_{\alpha \beta }$ is the induced metric
on the spacelike surfaces defined by $t=const$. We make some further
definitions:
\begin{eqnarray}
E_{\alpha }^{i} &=&h_{\alpha }^{\beta }e_{\beta }^{i}, \\
A_{a}^{i} &=&h_{\alpha }^{\beta }{}^{3}A_{\beta }^{i}, \\
B_{a} &=&h_{\alpha }^{\beta }{}^{3}A_{\beta \hat{0}}, \\
F_{\alpha \beta I} &=&h_{\alpha }^{\gamma }h_{\beta }^{\delta
}{}^{3}F_{\gamma \delta I}, \\
(n\cdot A)^{I} &=&n^{a}{}^{3}A_{a}^{I}.
\end{eqnarray}%
We can also use the gauge freedom in the definition of ${}^{3}A_{a}^{I}$ to
set $(n\cdot A)^{\hat{0}}=0$, and do so. We are still left with the freedom
to make $SO(2)$ rotations on the internal indices, $i=1,2$, with the angle
of rotation an arbitrary function of the spatial coordinates, ($x,y$).

\subsection{Field Equations}

In 2+1 spacetime dimensions:
\begin{eqnarray}
R_{abcd} &=&R_{ac}g_{bd}-R_{ad}g_{bc}+R_{bd}g_{ac}-R_{bc}g_{ad}+\left( \frac{%
R}{2}\right) \left( g_{ad}g_{bc}-g_{ac}g_{bd}\right) , \\
R_{ab} &=&\kappa (T_{ab}-Tg_{ab}), \\
R &=&-2\kappa T.
\end{eqnarray}%
So, therefore:
\[
R_{abcd}=\kappa \left[ T_{ac}g_{bd}-T_{ad}g_{bc}+T_{bd}g_{ac}-T_{bc}g_{ad}+T%
\left( g_{ad}g_{bc}-g_{ac}g_{bd}\right) \right] ,
\]%
where the curvature ${}^{3}F_{abI}$ is related to the $R_{abcd}$ by
\[
{}^{3}F_{abI}=R_{abcd}e^{cJ}e^{dK}\epsilon _{IJK}.
\]%
We define: $T^{IJ}=e^{aI}e^{bJ}T_{ab}$, so the Einstein equations can then
be written as:
\begin{eqnarray}
{\cal D}_{[a}e_{b]I} &=&0,  \label{eqn1} \\
{}^{3}F_{abI} &=&\kappa \left[ 2e_{[a}^{L}e_{b]}^{K}T_{L}^{J}\epsilon
_{IJK}-Te_{a}^{J}e_{b}^{K}\epsilon _{IJK}\right] .  \label{eqn2}
\end{eqnarray}%
Note that these equations remain well-defined even if $det(e)=\sqrt{-g}=0$,
so in this form the field equations actually describe a theory that is more
general than general relativity. When $e_{a}^{I}$ is invertible, however,
these field equations are fully equivalent to 3-d general relativity. We are
concerned with the dynamics of perfect-fluid spacetimes and for the purposes
of this section will consider only dust models, $p=0$, so $T^{IJ}=\kappa
\rho u^{I}u^{J}-\Lambda \eta ^{IJ}$, where $u_{K}u^{K}=-1$. We write $%
u^{k}=U^{k}=U_{k}$ and $u_{\hat{0}}=1+U_{k}U^{k}$, $k=1,2$.

\subsection{2+1 decomposition of field equations}

We now use $n^{a}$ and $h_{b}^{a}$ to decompose the field equations. From
equation (\ref{eqn1}) we find that $n\cdot A^{i}=0$ and:
\begin{eqnarray}
&\dot{E}_{\alpha i}=\epsilon _{ij}A_{\alpha }^{j},&  \label{edot1} \\
&\tilde{\eta}^{\alpha \beta }\partial _{\alpha }E_{\beta i}+\epsilon
_{ij}B_{\alpha }E_{\beta }^{j}\tilde{\eta}^{\alpha \beta }=0,&  \label{curv}
\\
&\tilde{\eta}^{\alpha \beta }\epsilon _{ij}A_{\alpha }^{i}E_{\beta }^{j}=0,&
\label{edot2}
\end{eqnarray}%
and by projecting equation (\ref{eqn2}) we arrive at:
\begin{eqnarray}
\dot{B}_{\alpha } &=&\kappa \rho u_{\hat{0}}U^{i}\epsilon _{ij}E_{\alpha
}^{j},  \label{Bdot} \\
\dot{A}_{\beta i} &=&\kappa \rho \left[ U_{k}U_{k}\epsilon _{ij}E_{\beta
}^{j}-U_{l}E_{\beta }^{l}\epsilon _{ij}U^{j}\right] -\Lambda \epsilon
_{ij}E_{\beta }^{j},  \label{Adot1} \\
\tilde{\eta}^{\alpha \beta }\partial _{\alpha }A_{\beta i}+\epsilon
_{ij}B_{\alpha }A_{\beta }^{j}\tilde{\eta}^{\alpha \beta } &=&-\kappa \rho
u_{\hat{0}}U^{j}\epsilon _{jk}E_{\alpha }^{k}E_{\beta i}\tilde{\eta}^{\alpha
\beta },  \label{curv2} \\
\tilde{\eta}^{\alpha \beta }(2\partial _{\alpha }B_{\beta }+\epsilon
_{ij}A_{\alpha }^{i}A_{\beta }^{j}) &=&\kappa \rho \left[ 2E_{\alpha
}^{k}E_{\beta }^{j}\tilde{\eta}^{\alpha \beta }U_{k}U^{i}\epsilon
_{ij}+E_{\alpha }^{i}E_{\beta }^{j}\tilde{\eta}^{\alpha \beta }\epsilon _{ij}%
\right] +\Lambda \epsilon _{ij}E_{\alpha }^{i}E_{\beta }^{j}\tilde{\eta}%
^{\alpha \beta }.  \label{epeqn}
\end{eqnarray}%
Equations (\ref{edot1}), (\ref{Bdot}) and (\ref{Adot1}) provide 10 evolution
equations for the ten degrees of freedom contained in the $E_{\alpha }^{i}$,
$B_{\alpha }$ and $A_{\alpha }^{i}$. To solve these we must specify these
quantities, as well as $\rho $ and the $U^{i}$, on some initial spacelike
hypersurface: a total of 13 initial functions. The other equations amount to
6 consistency conditions, which restricts the number of free functions to 7.
We may still make 3 coordinate transforms, and 1 internal $SO(2)$ rotation,
or gauge transform. This brings us down to $3$ functions that may be freely
specified on the initial surface; we take these to be $\rho $, $U^{1}$ and $%
U^{2}$. It is easy to check the the consistency conditions are preserved by
the evolution equations.

\section{General Cosmological Solutions with Comoving Dust}

The equations are particularly simple when the velocity can be chosen to be
co-moving (which, for 2+1 dimensional dust, is equivalent to it being
irrotational). We fix our coordinate system by choosing both $h_{\alpha
\beta }$ and $\dot{h}_{\alpha \beta }$ to be diagonal on our initial
surface; for a proof that this may always be done see \cite{Kriele}. It was
also proved by Kriele that if, initially, $\dot{h}_{\alpha \beta }\neq
\lambda (x,y)h_{\alpha \beta }$ then this coordinate choice is unique and
the only remaining freedom is $t\rightarrow t+t_{0}$, $x\rightarrow X(x)$
and $y\rightarrow Y(y)$ although we also retain the freedom to interchange $%
x $ and $y$. In this case, and only this case, we may use the coordinate and
gauge freedoms to move to a frame where the fluid is comoving and $E_{a}^{i}$
and $\dot{E}_{a}^{i}$ are diagonal at some initial instance, and then the
evolution equations ensure that they remain diagonal at all times. Such a
coordinate transform leads to eqn. (\ref{edot2}) being {\em automatically
satisfied}. Thus, in making this coordinate choice, we reduce the number of
consistency equations by $1$ and so the number of free-functions that can be
specified on some initial surface increases by $1$. We have already set $%
U^{1}=U^{2}=0$ so we are therefore left with 2 free functions. Solutions of
this system represent all $2+1$ dimensional dust spacetimes with vanishing
vorticity.

If $\dot{h}_{\alpha \beta }(t_{0})=\lambda (x,y)h_{\alpha \beta }(t_{0})$
for some $\lambda (x,y),$ then we may further transform to a conformal gauge
so that

\[
h_{\alpha \beta }=(1+\lambda (x,y)(t-t_{0}))e^{2\phi (x,y)}\delta _{\alpha
\beta },
\]

where $\delta _{\alpha \beta }={\rm diag}(1,1)$; such a coordinate choice is
unique up to $x\rightarrow Ax+B$, $y\rightarrow Ay+C$, $t\rightarrow t+t_{0}$%
. We keep the freedom to interchange $x$ and $y$. We shall see that such
spacetimes are specified by only {\em one} free function of $x$ and $y$.

We use our remaining coordinate freedom to set $h_{\alpha \beta }$ and $\dot{%
h}_{\alpha \beta }$ to be diagonal initially, and fix the gauge of our
internal index by requiring $E_{\alpha }^{i}$ and $\epsilon _{ij}A_{\alpha
}^{j}=\dot{E}_{\alpha i}$ to be diagonal initially. For comoving systems $%
U^{i}=0$, and so by combining equations (\ref{edot1}) and (\ref{Adot1}) we
see that
\[
\ddot{E}_{\alpha i}=\Lambda E_{\alpha i}.
\]%
Thus, if $E_{\alpha }^{i}$ and $\epsilon _{ij}A_{\alpha }^{j}=\dot{E}%
_{\alpha i}$ are diagonalised initially, they will remain diagonal for all
time. We have now fixed our system and our internal gauge freedom. With
these choices equation (\ref{edot2}) is automatically satisfied. Solving for
$E_{\alpha }^{i}$ gives us
\begin{eqnarray}
E_{1}^{\hat{1}} &=&C(x,y){\rm sh}_{\Lambda }(t)+D(x,y){\rm ch}_{\Lambda }(t),
\\
E_{2}^{\hat{2}} &=&W(x,y){\rm sh}_{\Lambda }(t)+V(x,y){\rm ch}_{\Lambda }(t),
\\
A_{1}^{\hat{2}} &=&C(x,y){\rm ch}_{\Lambda }(t)+\Lambda D(x,y){\rm sh}%
_{\Lambda }(t), \\
A_{2}^{\hat{1}} &=&-W(x,y){\rm ch}_{\Lambda }(t)-\Lambda V(x,y){\rm sh}%
_{\Lambda }(t).
\end{eqnarray}%
with $C(x,y)$, $D(x,y)$, $W(x,y)$ and $V(x,y)$ to be determined by the
remaining equations. We have defined ${\rm sh}_{\Lambda }(t)=\Lambda
^{-1/2}\sinh (\Lambda ^{1/2}t)$, and ${\rm ch}_{\Lambda }(t)=\cosh (\Lambda
^{1/2}t)$. All other components of $A_{\alpha }^{i}$ and $E_{\alpha }^{i}$
equal to zero. For comoving systems we also have
\[
\dot{B}_{\alpha }=0.
\]%
This equation is automatically satisfied whenever the two consistency
conditions, (\ref{curv}) and (\ref{curv2}), hold. In our choice of
coordinates and with our gauge-fixing these equations read:
\begin{eqnarray}
\partial _{y}C(x,y) &=&B_{1}W(x,y),  \label{ccond1} \\
\partial _{y}D(x,y) &=&B_{1}V(x,y),  \label{ccond2} \\
\partial _{x}W(x,y) &=&-B_{2}C(x,y),  \label{ccond3} \\
\partial _{x}V(x,y) &=&-B_{2}D(x,y).  \label{ccond4}
\end{eqnarray}%
The 2-surface, $t=const$, is flat whenever $\tilde{\eta}^{\alpha \beta
}\partial _{\alpha }B_{\beta }=0$. The solutions of these equations divide
into 3 distinct classes.

\subsection{Class 1: $B_{1} = 0$ and / or $B_2=0$}

If one or other of the $B_{a}$ vanishes we can, without loss of generality,
by interchange of $x$ and $y$, choose $B_{1}=0$ and leave $B_{2}$ to be
freely specified (we are free to choose $B_{2}=0$ if we wish). Except when $%
B_{2}=B_{2}(x)$, these solutions will generically require the 2-surface
given by $t=const$ to be non-flat. The solution of equations (\ref{ccond1})-(%
\ref{ccond4}) is simple in this case. We find $C=C(x)$, $D=D(x)$ and:
\begin{eqnarray}
W(x,y) &=&\int_{x_{0}}^{x}{\rm d}\xi C(\xi )\frac{\partial L(\xi ,y)}{%
\partial \xi }+f(y), \\
V(x,y) &=&\int_{x_{0}}^{x}{\rm d}\xi D(\xi )\frac{\partial L(\xi ,y)}{%
\partial \xi }+g(y), \\
B_{2} &=&-\frac{\partial L(x,y)}{\partial x}.
\end{eqnarray}%
We determine the energy density of the dust, from eqn. (\ref{epeqn}), to be
\[
\kappa \rho =\frac{-\partial _{x}^{2}L(x,y)+(W(x,y){\rm ch}_{\Lambda
}(t)+\Lambda V(x,y){\rm ch}_{\Lambda }(t))(C(x){\rm ch}_{\Lambda
}(t)+\Lambda D(x){\rm sh}_{\Lambda }(t)}{(C(x){\rm sh}_{\Lambda }(t)+D(x)%
{\rm ch}_{\Lambda }(t))(W(x,y){\rm sh}_{\Lambda }(t)+V(x,y){\rm ch}_{\Lambda
}(t))}-\Lambda ,
\]%
and the metric is
\[
{\rm d}s^{2}=-{\rm d}t^{2}+(C(x){\rm sh}_{\Lambda }(t)+D(x){\rm ch}_{\Lambda
}(t))^{2}{\rm d}x^{2}+(W(x,y){\rm sh}_{\Lambda }(t)+V(x,y){\rm ch}_{\Lambda
}(t))^{2}{\rm d}y^{2}.
\]%
This is equivalent to the $\partial V_{0}/\partial y=0$ solution found by
Kriele, ~\cite{Kriele}. As stated above, the limit $B_{2}\rightarrow 0$ is
well-defined in this case.

\subsection{Class 2: $B_{1}\neq 0,B_{2}\neq 0$ and $C/W\neq D/V$}

It will be seen that Class 1 solutions do not emerge as the $B_{1}=0$ limit
of the solution for $B_{1}$ and $B_{2}$ non-zero. The solution of the latter
case is more complicated. The condition $C/W\neq D/V$ ensures that we do not
have $\dot{h}_{\alpha \beta }\propto h_{\alpha \beta }$, and so our
coordinate choice is unique (up to rescalings of $x$ and $y$).

As discussed above, solutions of this class are defined by 2 free functions
of $(x,y)$; to aid finding the solution we choose these to be $C(x,y)$ and $%
D(x,y)$. By combining equations (\ref{ccond1})-(\ref{ccond4}), we can solve
for $B_{1}$: We solve this equation for $B_{1}$:
\[
\ln B_{1}:=\phi (x,y)+\ln S(y)=\int_{x_{0}}^{x}{\rm d}\xi \frac{C\partial
_{\xi }\partial _{y}D-D\partial _{\xi }\partial _{y}C}{C\partial
_{y}D-D\partial _{y}C}+\ln S(y),
\]%
where $S(y)$ and $x_{0}$ are arbitrary. Equations (\ref{ccond1}) and (\ref%
{ccond2}) can now be used to specify $W$ and $V$:
\begin{eqnarray}
W &=&e^{-\phi }\partial _{y}C/S(y), \\
V &=&e^{-\phi }\partial _{y}D/S(y).
\end{eqnarray}%
Finally, we check that these forms of $C$, $D$, $W,$ and $V$ define a unique
$B_{2}$ via equations (\ref{ccond3}) and (\ref{ccond4}). We find that they
do, and that $B_{2}$ is given by
\[
B_{2}=\frac{e^{-\phi }}{S(y)}\left( \frac{\partial _{y}C\partial
_{x}\partial _{y}D-\partial _{y}D\partial _{x}\partial _{y}C}{C\partial
_{y}D-D\partial _{y}C}\right) .
\]%
For $D\neq 0,$ we may write these in a slightly more familiar form by
defining $D=e^{\nu (x,y)}$ and $C=F(x,y)e^{\nu (x,y)}$. We then have
\begin{eqnarray}
\phi  &=&\ln F_{,y}+\nu +\alpha , \\
\alpha (x,y) &:&=\int_{x_{0}}^{x}{\rm d}\xi \frac{F_{,\xi }\nu _{,y}}{F_{,y}}%
.
\end{eqnarray}%
Thus, $B_{1}=S(y)F_{,y}e^{\nu +\alpha }$ and
\begin{eqnarray*}
W &=&\frac{e^{-\nu }(Fe^{\nu })_{,y}e^{-\alpha }}{F_{,y}S(y)}, \\
V &=&\frac{\nu _{,y}e^{\alpha }}{F_{,y}S(y)},
\end{eqnarray*}%
and
\[
B_{2}=\frac{e^{-\alpha -\nu }(\nu _{,y})^{2}}{S(y)(F_{,y})^{2}}\partial
_{x}\left( F+\frac{F_{,y}}{\nu _{,y}}\right) .
\]%
The metric is therefore given by:
\[
{\rm d}s^{2}=-{\rm d}t^{2}+R^{2}(x,y,t)e^{2\nu (x,y)}{\rm d}x^{2}+\frac{%
e^{-2(\nu +\alpha )}((R(x,y,t)e^{\nu (x,y)})_{,y})^{2}}{%
S^{2}(y)F_{,y}^{2}(x,y)}{\rm d}y^{2},
\]%
where $R(x,y,t)={\rm ch}_{\Lambda }t+F(x,y){\rm sh}_{\Lambda }t$, and the
functions $F(x,y)$ and $\nu (x,y)$ are arbitrary. Generalised Szekeres-like
solutions emerge in the limit $F_{,x}=0$. The requirement $C/W\neq D/V$
translates to $e^{\nu }\neq H(x)(F(x,y)-1)$ for some $H(x)$. If this
requirement does not hold then, although our solution is still valid, the
coordinate system is not uniquely specified and we may transform to a
conformal frame in which the solution is simplified; we deem such solutions
to be of Class 3 and deal with them in the following subsection. The energy
density for the Class 2 solutions is now given by a rather complicated
expression:
\begin{equation}
\kappa \rho =\frac{E(x,y;\Lambda )}{R(R_{,y}+R\nu _{,y})},
\end{equation}%
where
\begin{eqnarray}
E(x,y,\Lambda ) &=&e^{-2\nu }\left[ \nu _{,xy}\nu _{,x}-\nu
_{,xxy}-F_{,x}/F_{,y}^{2}(\nu _{,x}\nu _{,y}^{2}F_{,y}+F_{,x}\nu
_{,y}^{3}-3\nu _{,xy}\nu _{,y}F_{,y})\right.  \\
&&-\left. F_{,xy}/F_{,y}^{2}(\nu _{,y}\nu _{,x}F_{,y}+3\nu
_{,y}^{2}F_{,x}+2F_{,xy}\nu _{,y}-2\nu _{,xy}F_{,y})\right.   \nonumber \\
&&\left. +F_{,xx}\nu _{,y}^{2}/F_{,y}+F_{,xxy}\nu _{,y}/F_{,y}\right] +\frac{%
1}{2}e^{-2\nu }\left( Ke^{2\nu }\right) _{,y}  \nonumber
\end{eqnarray}%
with $K(x,y,t)=\dot{R}^{2}-\Lambda R^{2}-e^{2\alpha }(SF_{,y})^{2}$.
Generalised Szekeres solutions \cite{szek, BSS} emerge in the limit $F_{,x}=0
$ and the 2+1 Szekeres solutions \cite{debnath} themselves emerge when $%
e^{-\nu }=A(y)x^{2}+2B(y)x+C(y)$.

\subsection{Class 3: $B_{1}\neq 0,B_{2}\neq 0$ and $C/W=D/V$}

In this case $\dot{h}_{\alpha \beta }\propto h_{\alpha \beta }$ and so we
may, as mentioned above, transform to a conformal gauge where both the
metric and its time derivative are proportional to $\delta _{ab}$. In this
case, without loss of generality, $D=V=e^{\phi (x,y)}$ and $C=W=\mu e^{\phi
(x,y)}$, $B_{1}=\partial _{y}\phi $ and $B_{2}=-\partial _{x}\phi $; $\phi $
is arbitrary and $\mu $ is a constant. The metric then becomes
\[
{\rm d}s^{2}=-{\rm d}t^{2}+e^{2\phi }(\mu {\rm sh}_{\Lambda }(t)+{\rm ch}%
_{\Lambda }(t))^{2}\left( {\rm d}x^{2}+{\rm d}y^{2}\right)
\]%
and the energy density is:
\[
\kappa \rho =\frac{-e^{-2\phi }\nabla ^{2}\phi (x,y)+(\mu {\rm ch}_{\Lambda
}(t)+\Lambda {\rm sh}_{\Lambda }(t))^{2}}{(\mu {\rm sh}_{\Lambda }(t)+{\rm ch%
}_{\Lambda }(t))^{2}}-\Lambda
\]%
where $\nabla ^{2}=\partial _{x}^{2}+\partial _{y}^{2}$. The FRW solution
emerges from the case where $e^{-2\phi }\nabla ^{2}\phi (x,y)={\rm const}$.
This is also the only homogeneous limit of this class of solutions.

\section{Cosmological Solutions with Non-Comoving Dust}

{\em \ }In this section we will extend our discussion to include the
remaining degrees of freedom in the cosmological evolution of dust. We
consider non-comoving fluid motions and seek a new class of non-comoving
solutions where only one of the spatial velocity components is non-zero;
without loss of generality, we will take this to be the $x$-component. We
will initially work with $\Lambda =0$, however we shall present a simple
transformation that allows us to map $\Lambda =0$ solutions into $\Lambda
\neq 0$ ones.

First, we use some of the remaining gauge and coordinate freedom to set $%
E_{2}^{\hat{1}}=0$. With this choice we have $U_{\hat{x}}=\sinh \theta \neq
0 $, $U_{\hat{2}}=0$ and $u_{\hat{0}}=\cosh \theta $. This is equivalent to
demanding that we choose our local frame field so that the $y$-component of
the fluid velocity vanishes. As before, we have that $\dot{E}_{\alpha
i}=\epsilon _{ij}A_{\alpha }^{j}$, and the relation $\tilde{\eta}^{\alpha
\beta }\epsilon _{ij}A_{\alpha }^{i}E_{\beta }^{j}=0$ tells us that $\dot{E}%
_{2}^{\hat{2}}/E_{2}^{\hat{2}}=\dot{E}_{1}^{\hat{2}}/E_{1}^{\hat{2}},$ which
implies $E_{1}^{\hat{2}}=A(x,y)E_{2}^{\hat{2}}$. Hence, we find that $E_{1}^{%
\hat{1}}=C(x,y)t+D(x,y)$. The remaining (independent) equations to be solved
now read:
\begin{eqnarray}
\ddot{E}_{2}^{\hat{2}} &=&-\kappa \rho \sinh ^{2}\theta E_{2}^{\hat{2}}, \\
\dot{B}_{2} &=&\kappa \rho \cosh \theta \sinh \theta E_{2}^{\hat{2}}, \\
B_{1} &=&A(x,y)B_{y}(x,y,t)+f(x,y), \\
\partial _{y}(C(x,y)t+D(x,y)) &=&f(x,y)E_{2}^{\hat{2}},  \label{solve1} \\
\left( \partial _{x}-A_{,y}-A\partial _{y}\right) E_{2}^{\hat{2}}
&=&-B_{2}(C(x,y)t+D(x,y)), \\
\kappa \rho \cosh ^{2}\theta (C(x,y)t+D(x,y))E_{2}^{\hat{2}} &=&\left(
\partial _{x}-A_{,y}-A\partial _{y}\right) B_{y}-\partial _{y}f(x,y) \\
&&+C(x,y)\dot{E}_{2}^{\hat{2}}.  \nonumber
\end{eqnarray}%
By taking two time derivatives of equation (\ref{solve1}), we can see that
we must have $f(x,y)=0$, which in turn implies $C=C(x)$ and $D=D(x)$. This
is therefore a generalisation of the Class 1 comoving solutions found in the
last section to case of non-zero vorticity. In the system of equations
above, we have written down two evolution equations and two consistency
equations, and by combining the evolution equation for $B_{\alpha }$ with
the time derivative of the first of the consistency equations we arrive at a
third condition. On any initial surface, we must specify $6$ functions of $x$
and $y$: $A(x,y)$, $\rho $, $\theta $, $E_{2}^{\hat{2}}$ and $\dot{E}_{2}^{%
\hat{2}}$, and $B_{2}$, and 2 functions of $x$: $C(x)$ and $D(x)$. Given the
assumed form of the metric, the only coordinate freedom we have left is $%
y\rightarrow Y(x,y)$, $x\rightarrow X(x)$ and $t\rightarrow t+t_{0}$. At
some initial instant we can always use the residual coordinate freedom to
set $A(x,y)=0$. Without loss of generality, therefore, we set $A(x,y)=0$.
This leaves $5$ free functions and $3$ consistency equations. In total, we
are left with 2 free functions that can be freely specified on the initial
surface, one less than required for the general solution of the Einstein
equations in accord with our setting one of the 2-velocity components equal
to zero.

We shall define new variables: $X=\kappa \rho \cosh ^{2}\theta $, $Y=\tanh
\theta $, $B_{2}=L(x,y,)$ and $E_{2}^{\hat{2}}=M(x,y,t)$. With these
definitions the equations become:
\begin{eqnarray}
\ddot{M} &=&-XY^{2}\,M,  \label{eqn1a} \\
\dot{L} &=&XY\,M,  \label{eqn2a} \\
\partial _{x}M &=&-L(C(x)t+D(x)),  \label{eqn3a} \\
X(C(x)t+D(x))M &=&\partial _{x}L+C(x)\dot{M}.  \label{eqn4a}
\end{eqnarray}%
When $C\neq 0$ we can, at least locally, set $C=1,$ without loss of
generality, by using the freedom to redefine the $x$ coordinate. In this
case, we can combine the above equations (\ref{eqn1a})-(\ref{eqn4a} )into a
single second-order, non-linear, PDE for $M(x,y,t)$:
\begin{equation}
\left( \partial _{t}\left( \frac{\partial _{x}M}{t+D}\right) \right)
^{2}(t+D)^{2}=\left( (t+D)\partial _{x}\left( \frac{\partial _{x}M}{t+D}%
\right) -(t+D)\partial _{t}M\right) \partial _{t}^{2}M.  \label{geneqn}
\end{equation}%
We note that via the coordinate transformation $(t,x) \rightarrow (t',x')$, defined below, we can, without loss of generality, set $D(x)=0$ and preserve all the assummed properties of the metric and matter content.
\begin{eqnarray*}
t'\cosh x' = t\cosh x + \int^{x} d\xi D(\xi)\sinh \xi, \\
t'\sinh x' = t\sinh x + \int^{x} d\xi D(\xi)\cosh \xi.
\end{eqnarray*}
Hereafter we fix our coordinate system by taking $D=0$.

\subsection{General solution for dust with one non-comoving velocity}
By the coordinate transform defined above we set $D=0$ and solve the resulting system of equations. The only subcase not explicitely covered by this solution will be that when $C=0$ (in which case we can take $D=1$ w.l.o.g.).    We shall see later through the class of solutions with $C\neq 0$ is, up to a coordinate transform, equivalent to the class of
solutions with $C=0$, $D=1$. We define a new variable $Z=XM$. The system of
equations now reads:
\begin{eqnarray*}
\ddot{M} &=&-ZY^{2}, \\
\dot{L} &=&ZY, \\
\partial _{x}M &=&-Lt, \\
tZ &=&\partial _{x}L+\dot{M}.
\end{eqnarray*}%
By combining the first three equations we find: $Y=u_{,\tau }/u_{,x}$ where $%
\tau =\ln t$ and $u=M_{,\tau }-M$. We then take the $\tau $-derivative of
the last equation to arrive at:
\[
\partial _{\tau }\Omega -Y\partial _{x}\Omega =\partial _{x}Y-Y^{2},
\]%
where we have defined $e^{\Omega }=Zt$. We now make a coordinate transform: $%
(x,y)\rightarrow (u,X)$, where $u=M_{,\tau }-M$ and $X=x$. With respect to
these new coordinates, the above equation becomes:
\[
\left( \Omega -\ln u_{,x}\right) _{,X}=Y
\]%
In terms of these new coordinates $1/Y=-\tau _{,X}$. We can rewrite $\ddot{M}%
=-ZY^{2}$ as $u_{,\tau }e^{-\tau }=-ZtY^{2}$. Inserting this into the above
equation leads to a simple equation for $Y$:
\[
Y_{,X}=Y^{2}-1,
\]%
which has solution $Y=\tanh (x-d(u,y))$ with $d(u,y)$ a function of
integration. Solving the above equations we then find
\[
\frac{e^{\Omega }}{u_{,x}}=\frac{Zt}{u_{,x}}=C(u,y)\cosh (x-d(u,y)),
\]%
with $C(u,y)$ a function of integration. From $u_{,\tau }e^{-\tau }=-ZtY^{2}$
we have
\[
e^{-\tau }=-C(u,y)\sinh (x-d(u,y)).
\]%
It will be more straightforward to define $F^{\prime
}(u,y):=F_{,u}=-C(u,y)\cosh d(u,y)$ and $G^{\prime
}(u,y):=G_{,u}=C(u,y)\sinh d(u,y)$. It is clear that $F^{\prime
}{}^{2}-G^{\prime }{}^{2}>0$. With respect to these definitions the above
equation for $\tau $ becomes:
\[
e^{-\tau }=\frac{1}{t}=F^{\prime }\sinh (x)+G^{\prime }\cosh (x).
\]%
From this, we find expressions for $u_{,\tau }$ and $u_{,x}$:
\begin{eqnarray*}
u_{,\tau } &=&-\frac{F^{\prime }\sinh (x)+G^{\prime }\cosh (x)}{F^{\prime
\prime }\sinh (x)+G^{\prime \prime }\cosh (x)}, \\
u_{,x} &=&-\frac{F^{\prime }\cosh (x)+G^{\prime }\sinh (x)}{F^{\prime \prime
}\sinh (x)+G^{\prime \prime }\cosh (x)}.
\end{eqnarray*}%
Finally, we find $M:=tP$ by solving $e^{\tau }\partial _{\tau }(e^{-\tau
}M)=u$ to obtain
\[
P=-u^{2}(\frac{F}{u})^{\prime }\sinh (x)-u^{2}(\frac{G}{u})^{\prime }\cosh
(x)+H(x,y)
\]%
where the function $H(x,y)$ is found, by insertion of $M$ into $tZ=\partial
_{x}L+\dot{M}$, to satisfy $H_{,xx}=H$. We can therefore absorb $H(x,y)$
into the definition of $F(u,y)$ and $G(u,y)$, and without loss of generality
set $H=0$. Thus, we have the final form of the general non-comoving dust
solution with $U_{y}=0:$
\begin{eqnarray}
M &=&-tu^{2}\left( \left( \frac{F}{u}\right) ^{\prime }\sinh x+\left( \frac{G%
}{u}\right) ^{\prime }\cosh x\right) , \\
\kappa \rho &=&\frac{F^{\prime }{}^{2}-G^{\prime }{}^{2}}{t^{2}\left(
F^{2}(u/F)^{\prime }\sinh x+G^{2}(u/G)^{\prime }\cosh x\right) \left(
F^{\prime \prime }\sinh x+G^{\prime \prime }\cosh x\right) }, \\
Y &=&\frac{F^{\prime }\sinh x+G^{\prime }\cosh x}{F^{\prime }\cosh
x+G^{\prime }\sinh x}, \\
U_{x} &=&\frac{{\rm sign}(Y)}{\sqrt{F^{\prime }{}^{2}-G^{\prime }{}^{2}}}, \\
U_{0} &=&\frac{\left\vert F^{\prime }\cosh x+G^{\prime }\sinh x\right\vert }{%
\sqrt{F^{\prime }{}^{2}-G^{\prime }{}^{2}}}.
\end{eqnarray}%
It is evident from the form of the density, $\rho ,$ that there is a
curvature singularity at $t=0$, and that to ensure we always have
positive energy densities we must have
\[
\left( F^{2}(u/F)^{\prime }\sinh x+G^{2}(u/G)^{\prime }\cosh x\right) \left(
F^{\prime \prime }\sinh x+G^{\prime \prime }\cosh x\right) \geq 0,
\]
with curvature singularities appearing in the case of equality. We could, it
should be noted, rewrite this as the requirement that
\[
\frac{(P^{2})^{\prime }}{u}=(u^{2}(F/u)^{\prime }\sinh x+u^{2}(G/u)^{\prime
}\cosh x)^{2\prime }/u\leq 0,
\]%
with singularities forming in the case of equality. In the next subsection
we shall consider the form and nature of these singularities in more detail.
Using the above results we can find the expansion scalar, $\Theta $, for
these spacetimes:
\[
\Theta =(-u)\frac{A(u,y)\sinh x-B(u,y)\cosh x}{tP(u,x,y)P^{\prime }}
\]%
where
\begin{eqnarray*}
A(u,y) &=&\left( 2FG^{\prime }F^{\prime \prime }-GF^{\prime }F^{\prime
\prime }-FF^{\prime }G^{\prime \prime }+F^{\prime 2}G^{\prime \prime
}u-G^{\prime }F^{\prime }F^{\prime \prime }u\right) , \\
B(u,y) &=&\left( 2GF^{\prime }G^{\prime \prime }-FG^{\prime }G^{\prime
\prime }-GG^{\prime }F^{\prime \prime }+G^{\prime 2}F^{\prime \prime
}u-G^{\prime }F^{\prime }G^{\prime \prime }u\right) .
\end{eqnarray*}%
We also find that the vorticity of these spacetime is:
\[
\omega ^{2}=\frac{\left( G_{,y}^{\prime }F^{\prime 2}-G^{\prime }F^{\prime
}F_{,y}^{\prime }\right) ^{2}-\left( G^{\prime 2}F_{,y}^{\prime }-F^{\prime
}G^{\prime }G_{,y}^{\prime }\right) ^{2}}{\left( F^{\prime 2}-G^{\prime
2}\right) ^{3}t^{2}P^{2}}.
\]

\subsection{Classification of Singularities}

In addition to the standard cosmological singularity we have singularities
in the dust density ($\rho \rightarrow \infty $) whenever:
\[
(P^{2})^{\prime }=0.
\]%
We can divide these $(P^{2})^{\prime }$ type singularities into two distinct
classes: class A singularities are where $P\neq 0$, $P^{\prime }=0$, and
class B are where $P=0$. Class B singularities can be thought of as
shell-focusing singularities, in analogy to those in the Szekeres
spacetimes. For class A singularities, the volume element of the metric
remains non-zero, whereas for class B singularities it vanishes. From the
definition of $u$, we have that $u=0$ iff $u_{,\tau }=0$ which will not be
the case at any finite time. Thus, at finite times, either $u>0$ or $u<0$.

Class A and B singularities are generically naked. This can be seen by
consider geodesics that move along $y=const$ paths. The metric along ${\rm d}%
y=0$ is:
\[
ds^{2}=-{\rm d}t^{2}+t^{2}{\rm d}x^{2}=-t^{2}{\rm d}u{\rm d}v,
\]%
where $u=\ln t+x$ and $v=\ln t-x$. From any point, $\{t_{0},x_{0},y_{0}\},$
we can therefore move along an outward-moving null geodesic, defined by $%
y=0,v=const$, that will reach null infinity. It follows that there exist no
black-hole horizons in this spacetime. We now consider the strength of a
singularity at the point $\{t_{0},x_{0},y_{0}\}=\{u_{0},x_{0},y_{0}\}$. We
consider the quantity
\[
\Psi :=R_{ab}K^{a}K^{b},
\]%
where $K^{a}={\rm d}x^{a}(k)/{\rm d}k$, and $k\in (0,1]$. The singularity
lies at $k=0$. Since the Weyl tensor vanishes in 2+1 dimensions, by
propositions 1-4 of Clarke and Krolak, \cite{krolak}, a necessary and
sufficient condition for the singularity to be {\it strong} in the sense of
Krolak is that the following integral does not converge as $k\rightarrow 0$:
\[
J(k)=\int_{1}^{k}{\rm d}k^{\prime }\Psi (k^{\prime }).
\]%
If $\lim_{k\rightarrow 0}J$ does not exist then the {\it limiting focusing
condition} (LFC) is said to apply. If $J(k)$ does converge as $k\rightarrow
0 $ then the singularity is {\it weak} in the sense of Krolak, and also in
the sense of Tipler \cite{tipler}. A necessary and sufficient condition for
the singularity to be {\it strong} in the sense of Tipler is that the strong
LFC apply, i.e. that $J(k)$ not be integrable in $(0,1]$. If $J(k)$ is
integrable in $(0,1]$ then the singularity is Tipler {\it weak}.

The equations describing null geodesics in this background are:
\begin{eqnarray}
K^{t} &=&\frac{{\cal P}}{t},\qquad K^{x}=\frac{{\cal Q}}{t^{2}},\qquad K^{y}=%
\frac{l}{t^{2}M^{2}}, \\
{\cal P}^{2} &=&{\cal Q}^{2}+\frac{l^{2}}{M^{2}}, \\
\frac{{\rm d}{\cal P}}{{\rm d}k} &=&\frac{ul^{2}}{t^{3}P^{3}}, \\
\frac{{\rm d}{\cal Q}}{{\rm d}k} &=&-\frac{(F^{\prime }\cosh x+G^{\prime
}\sinh x)l^{2}}{t^{2}P^{3}}, \\
\frac{{\rm d}l}{{\rm d}k} &=&-\frac{(F_{,y}\sinh x+G_{,y}\cosh x)l^{2}}{%
t^{2}P^{3}}.
\end{eqnarray}%
We can see that simple solutions can be found if we take $l=0=K^{y}$. In
these cases ${\cal P}=\pm {\cal Q}=const$. Therefore, along these null
geodesics we have
\[
t=\sqrt{t_{0}^{2}+2{\cal P}k},\qquad x=x_{0}\pm \tfrac{1}{2}\ln (1+2{\cal P}%
k/t_{0}^{2}).
\]%
Outward-moving null geodesics take the $+$ sign, while inward moving ones
correspond to the $-$ sign. We shall refer to these geodesics as `radial'
null geodesics (RNG). We define the quantity:
\[
\lambda =(F^{\prime }\cosh x+G^{\prime }\sinh x)(1\pm Y).
\]%
For finite $t>0$, $\lambda $ is both finite and non-zero. Along RNGs we
have:
\begin{eqnarray}
\frac{{\rm d}u}{{\rm d}k} &=&\mp \frac{{\cal P}\lambda u}{t^{2}P^{\prime }},
\\
\frac{{\rm d}P}{{\rm d}k} &=&\mp \frac{{\cal P}}{t^{2}}\left( \lambda
u-P_{,x}\right) ,  \label{pdiff} \\
\frac{{\rm d}P^{\prime }}{{\rm d}k} &=&\mp \frac{{\cal P}}{t^{2}P^{\prime }}%
\left( \lambda uP^{^{\prime \prime }}-P^{\prime }P_{,x}^{\prime }\right) .
\label{diffPp}
\end{eqnarray}

\noindent {\bf Proposition}: {{\em The LFC does not apply to RNGs
terminating on Class A singularities.} \newline
{\bf Proof}: Consider the quantity $\Psi $ for class $A$ singularities:
\[
\Psi =\frac{\lambda ^{2}(-u){\cal P}^{2}}{PP^{\prime }t^{4}}.
\]%
Using $1/P^{\prime }=\mp t^{2}/({\cal P}\lambda u)({\rm d}u/{\rm d}k)$ we
have:
\[
\Psi =\mp \frac{\lambda {\cal P}}{Pt^{2}}\frac{{\rm d}u}{{\rm d}k}.
\]%
The limit $\lim_{k\rightarrow 0}\lambda {\cal P}/Pt^{2}$ exists for class $A$
singularities and ${\rm d}u/{\rm d}k$ is integrable on $(0,\infty )$, and so
$\Psi $ is integrable on the same region. Therefore by propositions 4 and 6
of \cite{krolak} the LFC does not apply to RNGs for class A singularities.
Class A singularities are therefore gravitationally weak $\square $. \newline
\newline
}

\noindent {\bf Proposition}: {{\em The LFC applies along RNGs terminating on
Class B singularities provided $\lim_{k\rightarrow 0}1/P^{\prime }\neq 0$,
but the strong LFC does not apply.} \newline
{\bf Proof}: It is a sufficient condition for the LFC to hold that:
\[
\lim_{k\rightarrow 0}{k\Psi }=\lim_{k\rightarrow 0}\frac{k\lambda ^{2}(-u)%
{\cal P}^{2}}{PP^{\prime }t^{4}}\neq 0.
\]%
We note that, for $t\in (0,\infty )$:
\[
\lambda _{0}u_{0}-P_{,x}|_{0}=F(u_{0},y_{0})\cosh x_{0}+G(u_{0},y_{0})\sinh
x_{0}\pm u_{0}/t_{0}\neq 0,
\]%
and is finite. By l'H{\^{o}}pital's rule and equation (\ref{pdiff}), the
limit $\lim_{k\rightarrow 0}k/P$ must therefore exist and be non-zero. Thus $%
\lim_{k\rightarrow 0}{k\Psi }=0$ iff $\lim_{k\rightarrow 0}1/P^{\prime }=0$
or equivalently iff $\lim_{k\rightarrow 0}{\rm d}u/{\rm d}k=0$. The LFC
applies along all RNGs where $\lim_{k\rightarrow 0}{\rm d}u/{\rm d}k\neq 0$.
By propositions 1 and 2 of \cite{krolak} the strong LFC does not apply if
the integral
\[
J(k):=\int_{1}^{k}{\rm d}k^{\prime }\int_{1}^{k^{\prime }}{\rm d}k^{\prime
\prime }\Psi (k^{\prime \prime })
\]%
converges as $k\rightarrow 0$. Using integration by parts we see that this
is equivalent to the condition that $\int_{1}^{k}{\rm d}k^{\prime }k^{\prime
}\Psi (k^{\prime })$ converges. Using $1/P^{\prime }=\mp t^{2}/({\cal P}%
\lambda u)({\rm d}u/{\rm d}k)$ we have
\[
\Psi =\mp \frac{k}{P}\frac{\lambda {\cal P}}{t^{2}}\frac{{\rm d}u}{{\rm d}k}.
\]%
As shown above, the limit $\lim_{k\rightarrow 0}k/P$ exists and ${\rm d}u/%
{\rm d}k$ is integrable on $(0,1]$, therefore $k\Psi (k)$ is integrable on $%
(0,1]$ and the $J(k)$ converges, and the strong LFC does not apply $\square $%
.}

\subsection{Asymptotic Behaviour}

We consider next the asymptotics of our class of non-comoving dust
cosmologies and find the criteria for them to become homogeneous at late
times. We note that as $t\rightarrow \infty ,$ we have $F^{\prime }\sinh
x+G^{\prime }\cosh x\rightarrow 0^{+}$. We assume that as $t\rightarrow
\infty $, the quantity $F^{\prime \prime }\sinh x+G^{\prime \prime }\cosh x$
does not vanish. We also assume that the limit $\lim_{t\rightarrow \infty
}u=u_{0}$ exists and that all functions of $u$ have well-defined Taylor
series expansions about $u_{0}$. We must now consider two distinct cases:
the first where $\lim_{t\rightarrow \infty }(F^{\prime 2}-G^{\prime 2})\neq 0
$, and the second where this limit vanishes. In the first case we must have $%
U_{x}U^{x}\sim (F_{0}^{\prime 2}-G_{0}^{\prime 2})/t^{2}$ and so $%
\lim_{t\rightarrow \infty }Y=0$. At late times we therefore expect to
recover, at lowest order in $1/t$, a comoving spacetime. Physically, this
asymptotic time evolution just reflects simple momentum and angular momentum
conservation. An expanding area of radius $R,\ $velocity $U,$ and mass $%
M\propto \rho R^{2}$ will evolve so that $MVR$ is constant; hence $U\propto
\sqrt{U_{x}U^{x}}\propto R^{-1}\propto t^{-1}.$

If $\lim_{t\rightarrow \infty }(F^{\prime 2}-G^{\prime 2})=0$ then we have
that $F^{\prime }\sinh x+G^{\prime }\cosh x\rightarrow 0^{+}$ implies $%
F^{\prime }\cosh x+G^{\prime }\sinh x\rightarrow 0$ also; hence we must have
$\lim_{t\rightarrow \infty }F^{\prime }=\lim_{t\rightarrow \infty }G^{\prime
}=0$ and so $\lim_{t\rightarrow \infty}Y=\tanh (x+\theta (y))$ where $\tanh
(\theta )=\lim_{t\rightarrow \infty }G^{\prime }/F^{\prime }=G^{\prime
\prime }(u_{0},y)/F^{\prime \prime }(u_{0},y)$; this will be solvable for $%
\theta $ because of the requirement that $F^{\prime }{}^{2}>G^{\prime
}{}^{2} $ for all finite $t$. In this second case we should therefore expect
to recover an asymptotically comoving spacetime only approximately in some
region about $x+\theta (y)=0$.

Consider spacetimes where $\lim_{t\rightarrow \infty }(F^{\prime
2}-G^{\prime 2})\neq 0$; we find:
\[
t^{2}\kappa \rho \sim \frac{F_{0}^{\prime 2}-G_{0}^{\prime 2}}{\eta
_{0}(x,y)\lambda _{0}(x,y)}\left( 1+\frac{u}{t\eta _{0}}-\frac{\nu _{0}}{%
t\lambda _{0}^{2}}+\frac{2\sigma _{0}}{\lambda _{0}t}+{\cal O}(t^{2})\right)
\]%
where the subscript $0$ means that a quantity is evaluated at $u=u_{0}(x,y),$
and where:
\begin{eqnarray*}
\eta _{0}(x,y) &=&F_{0}(x,y)\sinh x+G_{0}(x,y)\cosh x, \\
\lambda _{0}(x,y) &=&F_{0}^{\prime \prime }(x,y)\sinh x+G_{0}^{\prime \prime
}(x,y)\cosh x, \\
\nu _{0}(x,y) &=&F_{0}^{\prime \prime \prime }(x,y)\sinh x+G_{0}^{\prime
\prime \prime }(y)\cosh x, \\
\sigma _{0}(y) &=&\frac{F_{0}^{^{\prime \prime }}(x,y)F_{0}^{\prime
}(x,y)-G_{0}^{^{\prime \prime }}(x,y)G_{0}^{\prime }(x,y)}{F_{0}^{\prime
2}(x,y)-G_{0}^{\prime 2}(x,y)}.
\end{eqnarray*}%
We can see that for such a spacetime to tend to homogeneity at late time, we
need
\[
\partial _{\alpha }\frac{F_{0}^{\prime 2}(x,y)-G_{0}^{\prime 2}(x,y)}{\eta
_{0}(x,y)\lambda _{0}(x,y)}=0,
\]%
where $\alpha $ stands for $x$ or $y$. We shall give an example of a class
of spacetimes where this condition holds below.

In the second of the two cases $F_{0}^{\prime 2}-G_{0}^{\prime 2}=0$, and
asymptotically we then find that either
\[
\kappa \rho \sim \frac{\cosh \theta (y)}{t^{4}\sinh ^{3}(x+\theta
(y))F_{0}^{\prime \prime }\eta _{0}(x,y)}+{\cal O}(t^{-5}),
\]%
if $\eta _{0}(x,y)\neq 0$, where $\eta _{0}$ is as defined above, or
otherwise
\[
\kappa \rho \sim \frac{\cosh \theta (y)}{t^{3}(-u_{0}(x,y))\sinh
^{3}(x+\theta (y))F_{0}^{\prime \prime }\eta _{0}(x,y)}+{\cal O}(t^{-4}),
\]%
if $\eta _{0}=0$ and $u_{0}\neq 0$. If $\eta _{0}=u_{0}=0$ then to leading
order the energy density is not positive at late times. We see that this
class of solutions does not have an isotropic and homogeneous FRW limit. We
can also see that to leading order it seems that we cannot avoid having a
timelike singularity at $x=-\theta (y)$ at late times in this class of
solutions. The early-time behaviour of both classes of solutions depends
strongly on the choice of the free functions $F(u,y)$ and $G(u,y)$. We shall
consider two examples which clearly illustrate the different extremes of
behaviour that are possible.

\subsubsection{Example 1}

Let us choose $F(u,y)=H(y)\sinh (u-s(y))$ and $G(u,y)=H(y)\cosh (u-s(y))$.
We have that:
\begin{eqnarray*}
\frac{1}{t} &=&H(y)\sinh (x+s(y)-u), \\
P &=&\cosh (x+u-s(y))-u\sinh (x+u-s(y)) \\
&=&\frac{1}{t}\left( \sqrt{1+H^{2}t^{2}}-\sinh ^{-1}(1/Ht)+x+s(y)\right) , \\
\kappa \rho  &=&\frac{H^{2}(y)}{t^{2}\cosh (x+s(y)-u)P(t,x,y)} \\
&=&\frac{H^{2}(y)}{t^{2}P(t,x,y)}\left( 1+\frac{1}{t^{2}H^{2}}\right)
^{-1/2}, \\
U_{x}U^{x} &=&1/H^{2}t^{2}.
\end{eqnarray*}%
By the analysis of the previous section, this spacetime has Krolak strong
singularities at $x=x_{0}(y,t)$, where $x_{0}(y,t)=\sinh ^{-1}\frac{1}{Ht}-%
\sqrt{1+H^{2}t^{2}}-s(y)$. We note that if $H>0$ then $\kappa \rho >0$ for
all $x>x_{0}(y,t)$; if $H<0$ then $\kappa \rho >0$ in the region $%
x<x_{0}(y,t)$; in either case we must restrict ourselves only to the region
where $\kappa \rho $ is positive. Since $P$ vanishes at $x=x_{0}$, the
circumference of the line defined by $x=x_{0}$, $t=0$ vanishes and this
should be properly considered to be a single point in the $x-y$ plane. This
is a shell-focusing singularity. We now consider the late and early-time
behaviour of this spacetime. \newline
\newline
{\bf {Early-time behaviour}}\newline
`Early' time now means $t<<1/H$, and we see that
\[
P(x,y,t)\sim t^{-1}\left( 1+\ln Ht/2+x+s(y)+{\cal O}((Ht)^{2})\right)
\]%
and
\[
\kappa \rho \sim \frac{H(y)}{\left( 1+\ln Ht/2+x+s(y)\right) }\left( (1+%
{\cal O}(H^{2}t^{2})\right) .
\]%
We can see from this expression that we can see that we will not be able to
reach the point $t=0$ since this leading order term will be become singular
at $t=t_{0}(x,y)>0$ where:
\[
t_{0}=(2/H)\exp \left( -x-s(y)-1\right) >0.
\]%
We should therefore interpret $t=t_{0}(x,y)$ as being the true initial
singularity - interestingly, we note that depending on our choice of $s(y)$,
the slice $t=t_{0}(x,y)$ can be either spacelike or timelike. As $%
t\rightarrow t_{0}$, $\rho $ diverges as $(t-t_{0})^{-1}$, which is weaker
than the we would otherwise expect had the fluid not been rotating (i.e. as $%
1/(t-t_{0})^{2}$). This singularity is of $P=0$ type, and so is Krolak
strong and Tipler weak. \newline
\newline
{\bf {Late-time behaviour}} \newline
`Late' time is $H^{2}t^{2}\gg 1$. We can see that
\[
P(x,y,t)\sim 1+(x+s(y))/Ht-\frac{1}{2}(Ht)^{-2}+{\cal O}((Ht)^{-3})
\]%
and:
\[
\kappa \rho \sim \frac{1}{t^{2}(1+(x+s(y))/Ht)}\left( 1+{\cal O}%
((Ht)^{-3})\right) ,
\]%
At late times (for fixed $x$ and $y$), this subclass of spacetimes tends to
a FRW limit:
\[
\kappa \rho \sim \frac{1}{t^{2}}.
\]

\subsubsection{Example 2}

A second class of illustrative spacetimes can we found by taking $%
F(u,y)=C(u)\cosh (\theta (y))$ and $G(u,y)=C(u)\sinh (\theta (y))$. At late
times, $C^{\prime }(u)\rightarrow 0$ and so these solutions fall into the $%
F_{0}^{\prime 2}-G_{0}^{\prime 2}=0$ class mentioned above. The reciprocal
time$\ $is given by
\[
\frac{1}{t}=C^{\prime }(u)\sinh (x+\theta (y)),
\]%
and
\[
P=(C(u)-uC^{\prime }(u))\sinh (x+\theta (y))
\]%
It is also straightforward to check that the expansion scalar vanishes for
this example, $\Theta =0,$ and so we must have the shear equal to the
vorticity: $\sigma ^{2}=\omega ^{2}$. Let us take, as an example, $%
C(u)=A_{0}+(C_{0}-A_{0})\cosh (u-u_{0})$, so $\sinh (u-u_{0})={\rm cosec}%
(x+\theta (y))/(C_{0}-A_{0})t$. We define the quantity $V:=(C_{0}-A_{0})t%
\sinh (x+\theta )$ and determine the early and late-time asymptotic
behaviours.\newline
\newline
{\bf {Early-time behaviour}}\newline
At early times, that is $|V|\ll 1$ for fixed $x$ and $y$, we have
\[
u\sim -\ln |V/2|+u_{0}+{\cal O}(V^{2})
\]%
and
\[
P\sim (A_{0}+\tfrac{1}{2})\sinh (x+\theta (y))+\frac{1}{t}\ln
|V/2|+(1-u_{0})/t+{\cal O}(V).
\]%
At early times, we there find that the energy density behaves as:
\[
\kappa \rho =\frac{1}{t^{2}\sinh ^{2}(x+\theta (y))}\left[ \left( \ln
|V/2|-(1-u_{0})+t(2A_{0}+1)\sinh (x+\theta (y))+{\cal O}(V^{2})\right) %
\right] ^{-1}.
\]%
Similarly to the previous example is seems as it we will not reach an the
point $t=0$, as we will encountered a singularity at $t=t_{0}>0$ where $%
t_{0}>0$ is the value of $t$ for which the quantity inside $[..]$ in
the above equation vanishes. This singularity is of $P=0$ type and
so is Krolak stronger and Tipler weak. As $t\rightarrow t_{0}$ we
will again have $\rho \propto (t-t_{0})^{-1}$. As in example one the
effect of rotation has been to weaken the strength of the initial
singularity. It is important to note that $t=t_{0}(x,y)$ is not
necessarily spacelike. In section 7.5.2 we shall describe an example
where there is no initial singularity but only a timelike $P=0$ type
`central' singularity.\newline

\noindent{\bf {Late-time behaviour}}\newline At late times, $|V|\gg
1$ we have: $u\sim u_{0}+V^{-1}+{\cal O}(1/V^{2})$ and $C(u)\sim
C_{0}+V^{-2}(C_{0}-A_{0})/2+{\cal O}(1/V^{4})$. Thus, we have
\[
P\sim C_{0}\sinh (x+\theta (y))-u_{0}/t-(1/2Vt)+{\cal O}(1/V^{3}).
\]%
We find for the energy density, assuming $C_{0}\neq 0$ that
\[
\kappa \rho \sim \frac{1}{t^{4}(C_{0}-A_{0})C_{0}\sinh ^{3}(x+\theta )}%
(1+u_{0}/(C_{0}\sinh (x+\theta )t))+{\cal O}(t^{-5}).
\]%
In the case $C_{0}=0,$ we have
\[
\kappa \rho \sim \frac{1}{t^{3}(C_{0}-A_{0})(-u_{0})\sinh ^{2}(x+\theta )}+%
{\cal O}(t^{-4}).
\]

\subsection{A non-vanishing cosmological constant}

Upon the inclusion of a cosmological constant term into the case with one
non-zero, non-comoving velocity in the $x$-direction, we find $E_{x}^{\hat{x}%
}=C(x){\rm sh}_{\Lambda }(t)+D(x){\rm ch}_{\Lambda }(t)$ where ${\rm sh}%
_{\Lambda }(t)=\Lambda ^{-1/2}\sinh (\Lambda ^{1/2}t)$ and ${\rm ch}%
_{\Lambda }(t)=\cosh (\Lambda ^{1/2}t)$. As in the $\Lambda =0$ case, we can
set both $E_{x}^{\hat{y}}$ and $E_{y}^{\hat{x}}$ to zero. We shall define
our variables as before: $X=\kappa \rho \cosh ^{2}\theta $, $Y=\tanh \theta $%
, $B_{y}=L(x,y,)$ and $E_{y}^{\hat{y}}=M(x,y,t)$, $Z=XM$. With these
definitions the equations become:
\begin{eqnarray*}
\ddot{M}-\Lambda M &=&-ZY^{2} \\
\dot{L} &=&ZY, \\
\partial _{x}M &=&-L(C(x){\rm sh}_{\Lambda }(t)+D(x){\rm ch}_{\Lambda }(t)),
\\
Z(C(x){\rm sh}_{\Lambda }(t)+D(x){\rm ch}_{\Lambda }(t)) &=&\partial
_{x}L+\left( C(x){\rm ch}_{\Lambda }(t)+\Lambda D(x){\rm sh}_{\Lambda
}(t)\right) \dot{M} \\
&-&(C(x){\rm sh}_{\Lambda }(t)+D(x){\rm ch}_{\Lambda }(t))\Lambda M.
\end{eqnarray*}%
When $\Lambda \neq 0$, we cannot always set $D=0$ whenever $C \neq 0$ without loss of generality; indeed it should be noted that those cases where $\vert D\vert > \vert C\vert/\sqrt{\Lambda}$ are qualitatively different where the opposite is true.  The former case will generically exhibit a bounce rather than an initial singularity.  If $D_{,x}=0,$, however, then we can, without loss of generality, redefine our $x$ and
$t$ coordinates so that $C=1$ and $D=0$. We now make the redefinitions $T=%
{\rm sh}_{\Lambda }(t)/{\rm ch}_{\Lambda }(t)=\Lambda ^{-1/2}\tanh (\Lambda
^{1/2}t)$, $\tilde{M}=M/{\rm ch}_{\Lambda }(t)$, $\tilde{Z}=Z{\rm ch}%
_{\Lambda }(t)$, and $\tilde{Y}=Y{\rm ch}_{\Lambda }(t)$. The system $\{%
\tilde{M},L,\tilde{Z},\tilde{Y};T\}$ then satisfies the $\Lambda =0$
equations for $\{M,L,Z,Y;T\}$.

\subsubsection{Generalisation of the $D_{,x}=0$ solution to $\Lambda \neq 0$}

We found the general solution for the $D=0$ and $\Lambda =0$ case
above. We have just seen that such solutions can be easily transformed into $%
\Lambda \neq 0$ solutions. We have
\[
M={\rm sh}_{\Lambda }(t)P(u,x,y),
\]%
where, as before,
\[
P(u,x,y)=-u^{2}\left( \left( \frac{F}{u}\right) ^{\prime }\sinh x+\left(
\frac{G}{u}\right) ^{\prime }\cosh x\right) .
\]%
The definition of $u$ in terms of $\{t,x,y\}$ changes to:
\[
\frac{1}{T}=\frac{{\rm ch}_{\Lambda }(t)}{{\rm sh}_{\Lambda }(t)}=F^{\prime
}(u,y)\sinh x+G^{\prime }(u,y)\cosh x.
\]%
The energy density is given by $\kappa \rho =Z(1-Y^{2})/M$. Using our
transformation,we have $\kappa \rho =\tilde{Z}(1-{\rm ch}_{\Lambda }^{-2}(t)%
\tilde{Y})/{\rm ch}_{\Lambda }^{2}(t)\tilde{M}$, and so
\begin{eqnarray}
\kappa \rho  &=&\frac{F^{\prime 2}-G^{\prime 2}+\Lambda }{{\rm ch}_{\Lambda
}^{2}(t)T^{2}\left( F^{2}(u/F)^{\prime }\sinh x+G^{2}(u/G)^{\prime }\cosh
x\right) \left( F^{\prime \prime }\sinh x+G^{\prime \prime }\cosh x\right) },
\\
Y &=&\frac{1}{{\rm ch}_{\Lambda }(t)}\frac{F^{\prime }\sinh x+G^{\prime
}\cosh x}{F^{\prime }\cosh x+G^{\prime }\sinh x}. \\
&&
\end{eqnarray}%
At late times $T\rightarrow \frac{1}{\sqrt{\Lambda }},$ and by applying the $%
\Lambda =0$ asymptotic analysis we can see that all solutions at late times
have energy densities that die off in time as $e^{-2\sqrt{\Lambda }t}$, and
in all cases $Y\rightarrow 0$ as $t\rightarrow \infty $. Thus, at late times
all such solutions become comoving iff $\Lambda >0$.

\subsection{New coordinates for $\Lambda =0$ case and nature of $t=0$}

When $\Lambda =0$ the metric of spacetime with $C=1$, $D=0$ is
\[
{\rm d}s^{2}=-{\rm d}t^{2}+t^{2}{\rm d}x^{2}+t^{2}P^{2}{\rm d}y^{2}.
\]%
The solutions are equivalent to all $C=0$, $D=1$, solutions under a
coordinate transform:
\[
t\rightarrow T=t\cosh x,\qquad x\rightarrow X=t\sinh x.
\]%
With these new coordinates it is easier to analyse what occurs when $t=0$.
The metric in $\{T,X\}$ coordinates is:
\[
{\rm d}s^{2}=-{\rm d}T^{2}+{\rm d}X^{2}+M^{2}{\rm d}y^{2},
\]%
where
\begin{eqnarray*}
M &=&F(u,y)X+G(u,y)T-u(T,X,y), \\
1 &=&F^{\prime }(u,y)X+G^{\prime }(u,y)T.
\end{eqnarray*}%
The energy density is given by
\[
\kappa \rho =\frac{F^{\prime }{}^{2}-G^{\prime }{}^{2}}{(FX+GT-u)(F^{\prime
\prime }X+G^{\prime \prime }T)}.
\]%
The line $t=0$ corresponds to $T=\pm X$, where $t\rightarrow 0$, $%
x\rightarrow x_{0}$, $x_{0}$ finite, is $T=X=0$, $X/Y=\tanh x_{0}$. This is
{\it not necessarily} a singularity.

Consider the behaviour near a point where $T=T_{0}=\pm X$ and $T_{0}$
finite; we have $1/T_{0}=\pm F^{\prime }(u_{0},y)+G^{\prime }(u_{0},y)$ and
so $\kappa \rho <\infty $ provided $\pm F(u_{0},y)+G(u+0,y)\neq u_{0}/T_{0}$
and $\pm F^{\prime \prime }(u_{0},y)+G^{\prime \prime }(u_{0},y)\neq 0$.
These conditions together simply require $\pm F(u,y)+G(u,y)\neq
u/T_{0}+o((u-u_{0})^{2})$ as $u\rightarrow u_{0}$. We also require $-\infty
<\mp F^{\prime }(u_{0},y)+G^{\prime }(u_{0},y)<\infty $. For most choices of
$F(u,y)$ and $G(u,y)$ the line $T=\pm X\neq 0$ will, therefore, be
non-singular. As we approaches $T=X=0,$ we must have that $G^{\prime
2}(u,y)\rightarrow \infty $, and since $F^{\prime 2}>G^{\prime 2}$ we
conclude that $F^{\prime 2}$ must also blow up at least as quickly. For many
choices of $F(u,y)$ and $G(u,y),$ we fill find that this corresponds to $%
u\rightarrow \pm \infty $, and so before we reach the point $T=X=0$ a $P=0$
type singularity is encountered, where $u=FX+GT$. {\em \ }

For now we assume that we can reach the point $X=T=0$ by moving along some
non-spacelike geodesic. We assume that at $X=T=0$, $u=u_{0}$ and that $u_{0}$
is finite. Near $u=u_{0}$ we assume that the blow up in $F^{\prime }$ and $%
G^{\prime }$ is due to a pole where $F(u,y)\sim A+C(u-u_{0})^{-m}$ and $%
G(u,y)\sim B+D(u-u_{0})^{n}$, with $m\geq n>-1$, $m,n\neq 0$, and $%
m^{2}C^{2}>n^{2}D^{2}\neq 0$. So, we have
\[
\kappa \rho \sim \frac{m^{2}C^{2}-n^{2}D^{2}(u-u_{0})^{2(m-n)}}{%
\Upsilon(X,T,u)}.
\]%
where:
\begin{eqnarray*}
\Upsilon(X,T,u)&=&(CX+D(u-u_{0})^{m-n}T+(AX+BT-u_{0})(u-u_{0})^{m}) \\
&&\cdot(m(m+1)CX+n(n+1)D(u-u_{0})^{m-n}T)
\end{eqnarray*}
If $m>0$ then the energy density behaves as:
\[
\kappa \rho \sim \frac{m^{2}C^{2}-n^{2}D^{2}(u-u_{0})^{2(m-n)}}{%
(CX+D(u-u_{0})^{m-n}T)(m(m+1)CX+n(n+1)D(u-u_{0})^{m-n}T)}
\]%
which clearly blows up as $T,X\rightarrow 0$. If $-1<m<0$ then $\kappa \rho
\propto (u-u_{0})^{m}X^{-1},$ which again is manifestly singular. Thus $%
T=X=0 $ is, in general, a curvature singularity. Similarly, if $u\rightarrow
\pm \infty $ as one approaches $T=X=0$ and $F(u,y)\sim A+Du^{m}$, $G\sim
A+Du^{n} $, $m\geq n\geq 1$, $m^{2}C^{2}\geq n^{2}D^{2}\neq 0$, then
\[
\kappa \rho \sim \frac{m^{2}C^{2}-n^{2}D^{2}u^{2(n-m)}}{%
(CX+Du^{n-m}T)(m(m-1)CX+n(n-1)Du^{n-m}T)},
\]%
which is manifestly singular at $X=T=0$. It is evident from these
asymptotics that the LFC applies along RNGs terminating at the singularity,
but the SFC does {\em not} apply.

As previously stated, it is often the case that the point $T=X=0$ is not
reachable since it lies behind a $P=0$ singularity. To illustrate this, and
to understand better the nature of these spacetimes we construct the Penrose
diagrams for two specific choices of $F(u,y)$ and $G(u,y)$.

\subsubsection{Example 1}

\begin{figure}[tbp]
\begin{center}
\epsfxsize=6cm \epsfbox{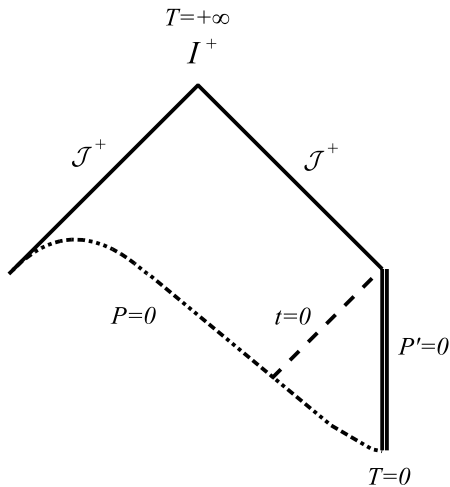}
\end{center}
\caption{Penrose Diagram for the spacetime in Example 1. We have take $y$
fixed and suppressed the $y$-axis. The singuarities are at $P=0$ and $%
P^{\prime }=0$. The line $t=0$ is also drawn.}
\label{fig1}
\end{figure}

We choose $F(u,y)=H(y)\sinh (u-s(y))$, $G(u,y)=-H(y)\cosh (u-s(y))$. This
gives:
\[
u=s(y)+\ln (H(y)(X+T))-\ln (1+\sqrt{1+H^{2}(y)(T^{2}-X^{2})}),
\]%
and
\[
\kappa \rho =H^{2}(y)\left[ \left( \sqrt{1+H^{2}(y)(T^{2}-X^{2})}%
+u(T,X,y)\right) \sqrt{1+H^{2}(y)(T^{2}-X^{2})}\right] ^{-1}.
\]%
There is a spacelike $P=0$, Krolak-strong and Tipler-weak singularity when
\[
s(y)+\ln (H(y)(X+T))-\ln (1+\sqrt{1+H^{2}(y)(T^{2}-X^{2})})+\sqrt{%
1+H^{2}(y)(T^{2}-X^{2})}=0,
\]%
and a timelike $P^{\prime }=0$, Krolak weak singularity at $%
X^{2}=1/H^{2}(y)+T^{2}$. The point $X=T=0$ lies beyond the boundary of this
spacetime. This space is homogeneous at late times: $\kappa \rho \sim 1/T^{2}
$, for fixed $y$ and $X$. We construct the Penrose diagram of this spacetime
for fixed $y$ in figure \ref{fig1}.
We can see from fig. \ref{fig1} that all past-directed timelike geodesics
terminate on a singularity, either on the `big-bang', singularity at $P=0$,
or on the timelike $P^{\prime }=0$ line. Although we have labelled the $P=0$
singularity as a `big-bang' it is important to note that it need not be
everywhere spacelike, and different choices of $H(y)$ and $s(y)$ can easily
result in it being timelike in some locality - that is appears spacelike in
fig. \ref{fig1} is due to our suppression of the $y$-axis. Since the
singularity at $P^{\prime }=0$ is weak in the sense of Krolak it is, in
principle, possible to continue through it, however we shall not consider
here what may lie beyond it.

\subsubsection{Example 2}

\begin{figure}[tbp]
\begin{center}
\epsfxsize=4cm \epsfbox{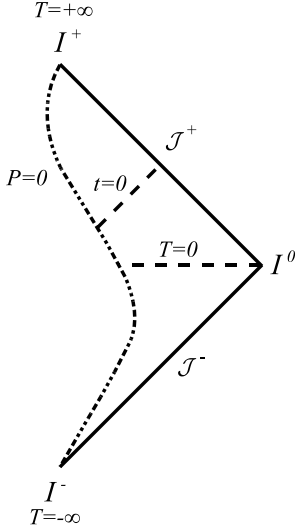}
\end{center}
\caption{Penrose Diagram for the spacetime in Example 2. We have take $y$
fixed and suppressed the $y$-axis. The only singularity is of $P=0$ type.
The lines $t=0$ and $T=0$ are also drawn.}
\label{fig2}
\end{figure}

We take $F=C(y)e^{u}\cosh (\theta (y))$, $G=C(y)e^{u}\sinh (\theta (y)),$ so
\[
u=-\ln (C(y)X\cosh (\theta (y))+C(y)T\sinh (\theta (y))).
\]%
In this case
\[
\kappa \rho =\frac{C^{2}(y)e^{2u}}{(1-u)},
\]%
and the only singularities are of $P=0$ type and occur when $u=1$. Unlike in
the previous example, the $P=0$ singularity is timelike in this case, and
can be thought of as a centre. At late times the space is not homogeneous
and $\kappa \rho \propto 1/(T^{2}\ln T)$. As before, the point $T=X=0$ lies
beyond the boundary of the spacetime. We construct the Penrose diagram for
this spacetime (at fixed $y$) in figure \ref{fig2}, from this it is clear
that there is no 'big-bang' initial singularity in this model. Indeed it can
be easily checked that the expansion scalar, $\Theta$, vanishes in for this
solution and so this is actually an example of an inhomogeneous static
spacetime with non-vanishing rotation and shear. Using the results of
sections 3.2 3.3 we see that we must have $\sigma^2 = \omega^2$ and ${\cal R}
= 2\kappa \rho$ where:
\[
\omega^2 = \frac{\theta_{,y}^2}{1-u}.
\]

\section{Scalar-Field Spacetimes}

\subsection{Solutions with one spacelike Killing vector}

We now find the general solution to 2+1 gravity with a massless scalar-field
source under the assumption that there is one spacelike Killing vector. The
metric takes the following form:
\[
{\rm d}s^{2}=-2A(u,v){\rm d}u{\rm d}v+C^{2}(u,v){\rm d}y^{2}.
\]%
We can transform this to $t,x$ coordinates by defining $u=(t+x)/\sqrt{2}$, $%
v=(t-x)/\sqrt{2}$. We assume a scalar field source, $\phi =\phi (u,v)$, with
energy momentum tensor: $T_{ab}=\partial _{a}\phi \partial _{b}\phi -\tfrac{1%
}{2}g_{ab}(\partial \phi )^{2}$ where $\phi $ satisfies the conservation
equation

\[
\square \phi =0.
\]

With these prescriptions the $(yu)$ and $(yv)$ components of Einstein
equations are satisfied trivially, and the $(yy)$ component requires $%
C_{,uv}=0$, the general solution of which is:
\[
C(u,v)=f(v)+g(u)
\]%
We also define $D(u,v)=f(v)-g(u)$, and move from $(u,v)$ to $(C,D)$
coordinates. By making different choices of $f(v)$ and $g(u)$ we can arrange
that either: $\partial _{a}C$ is spacelike and $\partial _{a}D$ is timelike,
or $\partial _{a}D$ is spacelike and $\partial _{a}C$ is time-like, or both $%
\partial _{a}C$ and $\partial _{a}D$ are null. In terms of $C$ and $D$ the $%
\phi $ wave equation reads:
\[
\phi _{,DD}=\phi _{,CC}+\phi _{,C}/C.
\]%
This is just the wave-equation in cylindrical polar coordinates with axial
and azimuthal symmetry (with $D$ playing the role of the usual time
coordinate and $C$ of the radial coordinate). We can solve this in terms of
Bessel functions.
\[
\phi =\int_{-\infty }^{\infty }{\rm d}kA(k)\left( \cos (\sqrt{k}D)+B(k)\sin (%
\sqrt{k}D)\right) \left( J_{0}(\sqrt{k}C)+E(k)Y_{0}(\sqrt{k}C)\right) ,
\]%
where $A(k),B(k),E(k)\in {\Bbb C}$ are arbitrary and $J_{0}$ and $Y_{0}$ are
zero-order Bessel functions of the first and second kind respectively.
Finally, we solve the equations for $A$ to give:
\[
\ln \left( \frac{A}{f^{\prime }(v)g^{\prime }(u)}\right)
=2C\int_{D_{0}(C)}^{D}{\rm d}D^{\prime }\phi _{,C}\phi _{,D}+F(C)
\]%
where
\[
F(C)=\int_{C_{0}}^{C}{\rm d}C^{\prime }C^{\prime }\left( \phi _{,C}^{2}+\phi
_{,D}^{2}\right) |_{D=D_{0}(C^{\prime })}.
\]%
Boundary conditions for $\phi $ are need to specify the solution further.

\subsection{PP-wave spacetimes}

In 2+1 spacetimes the metric for a scalar-field PP-wave spacetime can be
written in the form:
\[
{\rm d}s^{2}=H(u,x){\rm d}u^{2}+2{\rm d}u{\rm d}v+{\rm d}x^{2}.
\]%
The Einstein equations read $R_{uu}=-1/2H_{,xx}=\kappa (\phi _{,u})^{2}$ and
$R_{ab}=0$ otherwise; this implies $\phi =\phi (u)$, where $\phi (u)$ is
arbitrary. Solving for $H,$ we find:
\[
H=A(u)+B(u)x-\kappa (\phi ^{\prime }(u))^{2}~x^{2},
\]%
where $A(u)$ and $B(u)$ can both be freely specified. It can be checked that
in 2+1 dimensions these are the {\em only} perfect-fluid solutions (they are
equivalent to an irrotational $p=\rho $ fluid) that are compatible with the
PP-wave metric ansatz given above. Indeed, we find that the only permitted
choices of energy-momentum tensor must satisfy $T_{uu}\neq HT$ and $%
T_{ab}=Tg_{ab}$ otherwise. We find similar PP-wave solutions by considering
the 2+1 Einstein-Maxwell equations. Up to gauge transformations, all the
solutions are of the form $A_{u}=A_{v}=0$, and $A_{x}=\phi (u)$ for the
electromagnetic potential, with $\phi (u)$ arbitrary and $H(u,x)$ as given
above. The only non-vanishing components of $F_{ab}$ are $%
F_{ux}=-F_{xu}=\phi _{,u}(u).$

\section{Discussion}

%%%%%%%%%%%%%%%%%%%%
Whereas the general cosmological solutions of the 3+1 dimensional Einstein
equations are intractably complicated and likely dominated by
non-integrability, the structure of the theory in 2+1 offers the possibility
of making considerable progress towards finding the general solution in
several interesting situations. This fact, together with our current
perception that quantum field theory fits more naturally in three rather
than four dimensions, has motivated the study of Einstein's theory in
3-dimensional spacetimes.

In this article we employed covariant and first-order formalism techniques
to study the properties of general relativity in three dimensions. The
covariant approach provided an irreducible decomposition of the relativistic
equations and allowed for a mathematically compact and physically
transparent description of their properties. Using this information we
reviewed the kinematical, dynamical and geometrical features of
3-dimensional spacetimes and identified the special features that
distinguish them from the standard 3+1 models. These include the key role of
the isotropic pressure as the sole contributor to the gravitational mass of
the system and the fact that vorticity never increases with time. We also
reviewed the 3-d analogues of the spatially homogeneous and isotropic FRW
models and investigated their stability against linear perturbations. We
found that, unlike their conventional counterparts, dust-dominated 3-d
homogeneous and isotropic spacetimes are stable under shear and vorticity
distortions and (neutrally) stable against disturbances in their density
distribution. The latter reflects the vanishing of the total gravitational
mass in 3-dimensional dust models, which ensures the absence of linear
Jeans-type instabilities. In addition to isotropic spacetimes, we also
looked at 3-dimensional anisotropic models providing Kasner-like solutions
for the case of pressure-free matter and generalising G\"{o}del's universe
to three dimensions. The covariant formalism allowed us to carry out these
analyses by a study of the kinematic variables characterising the expansion
of the universe. The absence of both electric and magnetic Weyl curvature
components in three dimensions considerably simplifies the analysis. We then
specialised further to the case of a pressureless matter source. In addition
to being physically realistic, this assumption produces a significant
further simplification of the cosmological field equations in
three-dimensional spacetimes. We were able to find the general cosmological
solutions of the theory in the case where the matter was comoving. No
symmetry assumptions were made. We then considered the fully general
pressureless fluid system with non-comoving velocities. We were able to
solve the system in the case where one spatial velocity component was zero
whilst the other was non-zero. This allowed us to carry out an asymptotic
study, close to and far from singularities, of an inhomogeneous cosmology
with rotation, expansion and shear. All the singularities arising in these
solutions were classified using the different criteria of strength
introduced by Krolak and Tipler. We were able to provide a simple
transformation which generalised all the solutions we found with vanishing
cosmological constant into new solutions with non-zero cosmological
constant. Finally, we considered scalar-field metric with one Killing vector
and found all the PP-wave solutions in 2+1 dimensional universes.

These investigations suggestion a number of problems for further study.
Exact solutions in the cases with non-zero isotropic and anisotropic
pressure remain to be investigated. In the case of zero pressure, we have
analysed the problem of the general solution of the three-dimensional
Einstein equations into a well-defined system of partial differential
equations. We have solved for the case with comoving velocities and a single
non-comoving velocity but the problem remains to find the general solution
of the equations when both non-comoving fluid velocities are present.

\subsection*{Acknowledgement}

D. Shaw is supported by the PPARC.

\end{document}